%% file: main.tex
\documentclass[11pt]{article}
\usepackage{graphicx} 
\usepackage[utf8]{inputenc}
\usepackage[margin=1in]{geometry}
\usepackage{amsmath,amsthm,amssymb}
\usepackage[table]{xcolor}
\usepackage{xcolor}
\usepackage[hyphens]{url}
\usepackage{hyperref}
\usepackage[nameinlink,capitalise]{cleveref}
\usepackage{bbold}
\usepackage{mathtools}
\usepackage{doi}
\usepackage{tikz}
\usepackage{thm-restate}
\usepackage[ruled]{algorithm2e}
\usepackage[T1]{fontenc}
\usepackage{indentfirst}
\usepackage{libertine}
\usepackage{subcaption}
\usepackage{booktabs}

\usepackage{etoolbox}
\usepackage{todonotes}
\newtoggle{DEBUG}
\togglefalse{DEBUG}
\newtoggle{COMMENTS}
\toggletrue{COMMENTS}

\usetikzlibrary{arrows.meta, positioning}

\input{notations}

\title{Metric Distortion of Small-group Deliberation\footnote{All Python code is available at \url{http://bit.ly/3WHW8H4}}} 
\author{Ashish Goel\thanks{Management Science and Engineering Department, Stanford University, Stanford CA. Email: \texttt{\{ashishg,mohakg\}@stanford.edu}} \and Mohak Goyal\samethanks[1]  \and Kamesh Munagala\thanks{Department of Computer Science, Duke University, Durham, NC 27708-0129.  Supported by NSF awards CCF-2113798 and IIS-2402823. Email: \texttt{kamesh@cs.duke.edu}}}
\date{}

\begin{document}

\maketitle

\input{abstract}


\thispagestyle{empty}
\newpage
\setcounter{page}{1}
\input{introduction}

\input{model}

\input{copeland}

\input{average}

\input{average_two}

\input{average_three}

\input{average_sample}
\input{random}

\input{random_asymptotic}

\input{random_sample}

\input{random_general}
\input{conclusion}

\paragraph{Acknowledgment.} We thank Chris Bail and an anonymous STOC 2025 reviewer for carefully reading the paper and providing many illuminating comments. We have used ChatGPT4o to paraphrase some text, and to generate the Python code for our optimization problems. 

Subsequent to the publication of the conference version, we found that a new version of the global optimizer BARON solved our relaxation to $\theta_3$ to an improved and tight global optimum, which improved the upper bound in \cref{thm:theta3}. We also used interactive sessions with GPT5.5-Pro to revise the paper, specifically adding rigor to the statements and proofs of \cref{thm:lb_main,thm:sample1,thm:sample2}, and adding the rational certificate constructions in \cref{thm:random}. All AI generated content was carefully verified by the authors who take full responsibility for the work.

\newpage
\bibliographystyle{alpha}
\bibliography{refs}

\appendix

\end{document}

%% file: notations.tex
\newcommand{\E}{\mathbb{E}}

\newcommand{\alg}{\mathbf{Alg}}
\newcommand{\A}{\mathcal{A}}
\newcommand{\B}{\mathcal{B}}

\newtheorem{theorem}{Theorem}[section]
\newtheorem{lemma}[theorem]{Lemma}

\theoremstyle{definition}

\newtheorem{example}{Example}
\newtheorem{question}{Question}

\newcommand{\eps}{\varepsilon}

\crefname{Program}{Program}{Programs}
\creflabelformat{Program}{(#2\textup{#1})#3}

\renewcommand{\epsilon}{\varepsilon}

\makeatletter
\def\@fnsymbol#1{\ensuremath{\ifcase#1\or \dagger\or \ddagger\or
   \mathsection\or \mathparagraph\or \|\or **\or \dagger\dagger
   \or \ddagger\ddagger \else\@ctrerr\fi}}
\makeatother
\newcommand*\samethanks[1][\value{footnote}]{\footnotemark[#1]}

\definecolor{linkc}{rgb}{0.1, 0.5, 0.7}
\definecolor{citec}{rgb}{0.6, 0.3, 0.7}
\definecolor{urlc}{rgb}{0.5, 0.1, 0.2}
\hypersetup{
    colorlinks=true,
    linkcolor=linkc,
    citecolor=citec,
    urlcolor=urlc
}

%% file: abstract.tex
\begin{abstract}
We consider models for social choice where voters rank a set of choices (or alternatives) by deliberating in small groups of size at most $k$, and these outcomes are aggregated by a social choice rule to find the winning alternative. We ground these models in the metric distortion framework, where the voters and alternatives are embedded in a latent metric space, with closer alternative being more desirable for a voter. We posit that the outcome of a small-group interaction optimally uses the voters' collective knowledge of the metric, either deterministically or probabilistically. 

We characterize the distortion of our deliberation models for small $k$, showing that groups of size $k=3$ suffice to drive the distortion bound below the deterministic metric distortion lower bound of $3$, and  groups of size $4$ suffice to break the randomized lower bound of $2.11$. We also show nearly tight asymptotic distortion bounds in the group size, showing that for any constant $\epsilon > 0$, achieving a distortion of $1+\epsilon$ needs group size that only depends on $1/\epsilon$, and not the number of alternatives. We obtain these results via formulating a basic optimization problem in small deviations of the sum of $i.i.d.$ random variables, which we solve to global optimality via non-convex optimization. The resulting bounds may be of independent interest in probability theory.
\end{abstract}

%% file: introduction.tex
\section{Introduction}
Civic discourse is the cornerstone of modern democracy. A key challenge in modern social choice theory is how to leverage the power of discourse to 
improve societal decision making compared to simply voting on issues. Unfortunately, there are several roadblocks to this ideal. Much of the current online discourse tends to become acrimonious, increasing polarization rather than consensus. Despite some notable successes~\cite{Ingham18,fishkin1991democracy}, it is relatively cumbersome to organize large groups for agenda-driven discussions, and even when such discussions occur, they rarely result in meaningful opinion changes, leading to the same outcomes as direct voting. 

A promising approach emerging in social choice is small-group online forums, which are moderated and agenda-driven. Platforms such as the Online Deliberation Platform~\cite{stanfordd}, vTaiwan~\cite{vTaiwan}, Mismatch~\cite{mismatch}, and Myjunto~\cite{myjunto} have experimented with this approach. Recent work has shown the potential of AI mediation in these settings, including facilitating human interaction via summarization and rephrasing of diverse viewpoints~\cite{ChrisBail,ChatBot,stanfordd,LLM,ashkinaze2024}. 

Our goal is to model the outcome of such small group deliberations and analyze whether aggregating their outcomes is capable of finding a better societal alternative than would be found without such interaction. Our main result is that deliberation is strictly helpful even with small groups.

\subsection{Metric Distortion and Deliberation} 
We use the metric distortion framework~\cite{AnshelevichBEPS18} to analyze the quality of the deliberation. We assume there is one issue with several potential choices (called alternatives or candidates) that the voters are deliberating over. These voters and alternatives are embedded in a latent metric space. Voters reveal ordinal preferences (that is, a ranking) over alternatives that are consistent with the latent distances, where if an alternative is closer to the voter, it is more similar to the voter's opinion, hence ranked higher by the voter. 
We assume the set of alternatives is common to all voters, and a voter ranks all alternatives.

A social choice rule such as the Borda score or Copeland rule takes the rankings for all voters as input, and outputs an aggregate alternative. 
 Given an alternative, its $1$-median cost is the sum of the distances of all voters to this alternative. Consider all metrics consistent with voter rankings. The distortion is the worst case for all rankings and all metrics consistent with these rankings, of the ratio of the $1$-median cost of the alternative output by the social choice rule (that only knows the ordinal rankings) to the cost of the best alternative had the consistent metric been known. A lower distortion implies a better social choice rule, whose outcome is closer to the true $1$-median.

It is known that in the absence of voter interaction, any deterministic voting rule that uses only the ordinal preferences of the voters has distortion at least $3$~\cite{AnshelevichBEPS18} 
and any randomized rule has distortion at least $2.11$~\cite{AnshelevichP17,CharikarR22} 
 The key question we ask in this paper is:

\begin{question}
\label{q1}
Can small group deliberation improve these bounds?
\end{question}





To study this question formally,  we assume that voters deliberate in a small group and collectively reveal an ordinal preference (or ranking) that is consistent with their local view of the latent metric space. We consider several models of what ``consistent'' means, but a simple, canonical model is that the voters in the group could output the alternative that minimizes the sum of their distances to it (that is, their $1$-median alternative). The $1$-median intuitively captures persuasion -- consider two alternatives and two voters. If the first voter is very close to the first alternative, then they care strongly about that alternative. If the second voter is equidistant from both alternatives, they are indifferent. In this setting, the first voter can persuade the second to collectively output the first alternative, which is indeed the $1$-median. Even in the $1$-median setting, \cref{q1} is far from trivial. Indeed, the example below from~\cite{FainGMS17,CaragiannisM024} shows that even if all possible groups of voters of constant size output their $1$-median alternative, and these are aggregated by any social choice rule, the metric distortion remains $3$.


\begin{example}
\label{eg1}
Assume the group size is $k$, a constant. For some $\delta > 0$, there are $n$ voters, who are all at distance $1+\delta$ from a common candidate $c$. For each subset $S$ of voters of size $k$, there is a candidate $c_S$ that is at distance $1$ from the members of $S$, and distance $3$ from all other voters. Each group $S$ outputs $c_S$ when it deliberates, since this minimizes their $1$-median cost -- indeed, this alternative is preferred by all voters in $S$, leading to a trivial deliberation. Now any social choice method that aggregates the outcomes of these deliberations into a final winning alternative must produce a distribution over these $c_S$ as the final alternative. However, the average distance of all voters to $c$ is $1+\delta$, while for any $c_S$, the average distance is close to $3$. Therefore, the distortion is at least $3$.
\end{example}

To get around this lower bound, the group size needs to be super-constant, and indeed, the recent work of~\cite{CaragiannisM024}  shows that the $1$-median of a random sample of voters of size logarithmic in the number of candidates suffices to break the lower bound. However, such large groups lead to suboptimal group dynamics such as conformity, informational influence, and stereotyping of participants, which have been widely studied in psychology~\cite{Asch,Deutsch1955} and political science~\cite{Price}. In fact, the celebrated Asch experiments~\cite{Asch} show that conformity begins to kick in even with groups of around size $5$. Further, research in organizational psychology~\cite{axtell_2018} argues that groups of size larger than eight causes conversation quality to erode, people to become less candid, comments to become superficial, and tough topics avoided. The superiority of small groups has also been established via game-theoretic modeling of deliberation~\cite{Meirowitz}. All this means a large group interaction, even with the best intentions, is unlikely to find something as optimal as their $1$-median outcome, or allow participants to be ``public spirited''~\cite{FlaniganPW23} and compute the true social cost of each alternative. 
What is more likely is that the large group splits into multiple groups of constant size whose deliberation outcomes are aggregated by a moderator.  

\subsection{Deliberation Model with Small Groups}  
Although \cref{eg1} may seem to imply a negative result for \cref{q1} with groups of constant size, our main contribution is the following set of positive results.

\begin{quote}
There are natural models for multiple small group deliberations with constant-sized groups and natural social choice rules for aggregating their ordinal outcomes, which break the lower bounds of metric distortion (without deliberation) even with groups of size $k = 3$ or $4$, and where the distortion is $1 + O(1/\mbox{poly}(k))$ for larger group sizes $k$.
\end{quote}

For these results, we go beyond each deliberating group producing the $1$-median alternative. We do so by restricting the set of alternatives that a group can deliberate on, so that different small groups deliberate on different, focused sets of alternatives. In particular, we let each group deliberate on only two alternatives and output the better of the two. We then choose multiple groups of size $k$ at random from the population and let them deliberate between different pairs of alternatives, with the result of the deliberation being either the first or the second alternative. 
The outcome of these deliberations produces a tournament graph over the alternatives,  and we use the classic Copeland rule~\cite{voting-book} to aggregate this tournament into a single winning alternative. The entire model is presented in \cref{sec:prelim}.

For analyzing this framework, we assume that the outcome of any individual deliberation is a (possibly probabilistic) function of the strengths of the preference of voters within the group (as measured by their metric distances) for the two alternatives they are deliberating over. We consider two models: In the {\em averaging model}, the group deliberates between the two alternatives and outputs that alternative whose $1$-median cost is smaller.\footnote{In this model, instead of any group of size $k$ deliberating only over a pair of alternatives, we can equivalently let the group deliberate over all the alternatives and output a ranking based on their total distance to each alternative.}  In the {\em random choice model}, each group member has a normalized bias (or strength of preference) for one of the two alternatives proportional to the difference in their distance to the two alternatives. The probability with which an alternative is produced is proportional to the total normalized bias of voters in the group who prefer that alternative to the other. 
Note that in both cases, the output of a group is an ordinal preference.

\subsection{Our Results} We show analytic results for the distortion of both the averaging and random choice models as a function of the group size $k$, showing that a very small group size suffices to break the lower bound of metric distortion without deliberation. In particular, we show the following results, which are also summarized in \cref{tab:average,tab:random}.

\begin{table}[htp]
	\centering
 
	\begin{tabular}{|c|c|c|}
		\toprule
        Group size & Upper Bound & Lower Bound \\
        \midrule 
        $k = 2$ & $3 + \sqrt{2} \approx 4.414$ (\cref{thm:theta2}) & $1 + \sqrt{2} \approx 2.414$ (\cref{lem:theta2})  \\
        \midrule
        $k = 3$ & $2.78$ (\cref{thm:theta3}) & $\frac{5}{3} \approx 1.667$ (\cref{thm:lb1})  \\
           \midrule
        General $k$ & $1 + O\left(\frac{1}{k}\right)$ (\cref{thm:berry}) & $1 + \Omega\left(\frac{1}{k}\right)$ (\cref{thm:lb1}) \\
      \bottomrule
	\end{tabular}
	\caption{\label{tab:average} The upper and lower bounds on distortion for the Averaging model of deliberation. The upper bounds hold for the Copeland rule, while the lower bounds hold for any deterministic social choice rule that aggregates the ordinal rankings output by all groups of size $k$.  The upper bound of $3 + \sqrt{2} \approx 4.414$ is tight for uncovered set rules  when $k=2$. 
    }
\end{table}

\begin{table}[htbp]
	\centering
	\begin{tabular}{|c|c|c|c|}
		\toprule
        Group size & Upper Bound  (\cref{thm:random}) & Deterministic LB  & Randomized LB   \\
        \midrule 
        $k = 2$ &  $3.35$ & $1.82$ & $1.41$  \\
        \midrule
        $k = 3$ &  $2.31$ & $1.51$ & $1.25$ \\
           \midrule
        $k = 4$ &   $1.91$  & $1.37$ & $1.18$ \\
      \bottomrule
	\end{tabular}
	\caption{ \label{tab:random} The upper and lower bounds on distortion for the Random Choice model of deliberation. The upper bounds hold for the Copeland rule. The final two columns respectively show the lower bounds for any deterministic and randomized social choice rule that aggregates the ordinal rankings output by all groups of size $k$. These follow from \cref{thm:lb_main}.  For general $k$, the upper bound is $1 + O\left(\sqrt{\frac{\log k}{k}}\right)$ (\cref{thm:asymp1}), and approaches $1$ as $k \rightarrow \infty$.
    }
\end{table}

\begin{itemize}
\item For the averaging model 
with group size $k=3$, the distortion bound is $2.78$ (\cref{thm:theta3}).
This bound is smaller than the deterministic metric distortion lower bound of $3$. (As we discuss in \cref{sec:prelim}, the fair comparison for the averaging model is deterministic social choice rules.) For the random choice model, the distortion is $1.91$ for $k=4$ (\cref{thm:random}), which is lower than the lower bound of randomized metric distortion of $2.11$. We also show that for $k=2$, the distortion is exactly $3 + \sqrt{2} \le 4.42$ for the averaging model (\cref{thm:theta2}) and at most $3.35$ in the random choice model. The proofs use interesting non-convex programming relaxations for which we computationally bound the global optimum.\footnote{All Python code is available at \url{http://bit.ly/3WHW8H4}}

\item Next, we consider the asymptotic behavior of distortion in the size of the group, showing that the distortion approaches $1$ rapidly as the size of the group increases. For the averaging model, we show that groups of size $k = \Theta(1/\epsilon)$ are necessary and sufficient to achieve distortion $1+\epsilon$, where $\epsilon > 0$ is a constant (\cref{thm:berry,thm:lb1}). This result is an interesting application of the Berry-Esseen inequality~\cite{feller1971introduction}. For the random choice model, we show an upper bound of $k = \tilde{O}(1/\epsilon^2)$ (\cref{thm:asymp1}). These bounds on group size are {\em independent} of the number of alternatives, a result that cannot be obtained had each group only output its favorite alternative among all $m$ alternatives, for example, as in~\cite{CaragiannisM024}. 

\item Our analytic results assume a continuum of users from which infinitely many groups of size $k$ can be sampled. We also consider ``sample complexity'', which is the number of randomly sampled groups needed to achieve an additive $\epsilon$ to the analytically computed distortion. For constant $k \ge 2$ and small enough $\epsilon > 0$, we show that the sample complexity is $O\left(\frac{\log m}{\epsilon^{2}}\right)$ for the averaging model (\cref{thm:sample2}), and $O\left(\frac{m \log m}{\epsilon^2}\right)$ for the random choice model (\cref{thm:sample1}). 
\item In \cref{sec:general}, we present generalizations of the random choice model that are motivated by how the change in voter opinion depends on bias. We show that our analysis smoothly extends to these more general models, where the bounds become worse (asymptotically converging to numbers larger than one) and closer to the distortion of random dictatorship as the opinion change becomes less dependent on the bias of the voter.
\end{itemize}


From a practical perspective, our protocols and associated distortion bounds show that multiple tiny groups of people deliberating about specific alternatives suffice to improve on straightforward voting over all alternatives. This result is robust to natural models of deliberation. Since the goal of the paper is to make a conceptual point that very small group sizes suffice, we have focused on the Copeland rule. However, some of our analysis (such as \cref{thm:theta2}) can be extended to weighted tournament rules~\cite{MunagalaW19,Kempe_2020}, with comparable guarantees.

\subsection{Technical Highlight: Small Deviation Bounds and Non-convex Programs} 
It is known that the distortion of the Copeland rule can be characterized by its behavior on at most $3$ alternatives~\cite{AnshelevichBEPS18}; most of our analysis uses its behavior on two alternatives. As one of our contributions, we distill the type of analysis in~\cite{AnshelevichBEPS18} as solving a mathematical program about distributions. At a high level, such a mathematical programming approach is akin to the approach in~\cite{CharikarR22}, though the details are very different. For the averaging model, the problem of bounding the distortion reduces to a very basic question in probability theory (see \cref{eq:opt_avg} in \cref{sec:avg}): 
\begin{question}
    Given $k \ge 1$, find the distribution $D_k$ supported on $[-1,1]$ with maximum $\E[D_k]$, subject to the constraint that for $Y_1, Y_2, \ldots, Y_k \overset{\text{i.i.d.}}{\sim} D_k$, we have $\mbox{Median}\left(\frac{\sum_{i=1}^k Y_i}{k}\right) \le 0$.  
\end{question} 

Here, $k$ is the group size. We call the optimum expectation computed above $\theta_k$. This captures the worst-case gap between the mean and median of the average of bounded {\em i.i.d.} random variables. We show that the distortion is both upper and lower bounded by simple increasing functions of $\theta_k$, so that tightly bounding $\theta_k$ yields good bounds on distortion. As $k \rightarrow \infty$, by the Central Limit Theorem, the average of $i.i.d.$ and bounded random variables converges to a Normal distribution, so that $\theta_k \rightarrow 0$. Our goal is to bound $\theta_k$ 
for {\em small} $k$, and bound the asymptotic rate at which $\theta_k$ decreases with $k$. Our main technical result in \cref{sec:avg} is summarized in the theorem below and captures the combination of the bounds in \cref{thm:berry,thm:lb1,lem:theta2,thm:theta3}. This theorem may be of independent interest in probability theory.

\begin{theorem} [Proved in \cref{sec:avg}.]
\label{thm:main}
We have the following bounds\footnote{The work of~\cite{AnshelevichBEPS18}, as well as a direct analysis implies $\theta_1 = 0.5$, and is achieved by $D_1 = $ Bernoulli$(1,1/2)$.} on $\theta_k$: 
$$\theta_2 = \sqrt{2} - 1; \qquad \theta_3 \approx 0.25; \qquad \mbox{and}  \qquad  \theta_k = \Theta\left(\frac{1}{k}\right).$$ 
\end{theorem}

Though the above problem is easy to solve for groups of size $k=1$, even with groups of size $k = 2$, computing $\theta_2$ involves a non-convex convolution constraint over a distribution with infinite support,  whose expectation we need to globally optimize. Our main contribution is techniques to reduce the size of these programs via near-tight non-convex relaxations, so that a state of the art non-linear optimizer (BARON~\cite{Sahinidis1996,KS18}) can globally optimize it. With groups of size $k = 2$, we use pipage rounding~\cite{Ageev2004} combined with a relaxation to reduce the support size of the optimal distribution. For $k=3$, we use a technique in~\cite{Feige} to replace the identical random variables $\{Y_i\}$ with nonidentical ones, each with small support.  For $k \ge 4$, the problem becomes challenging. Indeed, the complexity of our relaxation grows significantly even in going from $k=2$ to $k=3$ and it is an open question whether there are tractable approaches even for $k=4$ that show reasonably tight bounds.

Conceptually, our mathematical programs have constraints on tail bounds of sum of independent random variables. However, we are interested in the median deviating from the mean, which corresponds to {\em small deviations} as opposed to large deviations (since the median is at most one standard deviation from the mean~\cite{MU}). There is scant work on small deviation bounds, the exceptions being the elegant works of~\cite{Feige,BertsimasP} (see also~\cite{HeZZ10}), and the works bounding the location of the median for the sum of special types of distributions~\cite{Siegel,Binomial,Poisson}. It is well-known that moment-based methods such as Bernstein's inequality~\cite{Boucheron} are slack in this regime, and in our setting, even tighter third moment bounds~\cite{BertsimasP} only yield weak results. This necessitates our non-convex programming approach. 

To tightly bound the asymptotic dependence of $\theta_k$ on $k$, we use the Berry-Esseen inequality~\cite{feller1971introduction}, which bounds the deviation in the CDF of the average of {\em i.i.d.} random variables from a Normal distribution. This is a substantial improvement over the $O(1/\eps^2)$ bounds that are often seen in social choice (e.g.~\cite{CaragiannisM024}) and cannot be achieved with moment-based methods, since unlike those methods, the Berry-Esseen inequality {\em becomes tighter} with increasing variance of the underlying random variables. 

For the random choice model in \cref{sec:random}, we show that the constraints in the non-convex program have a certain convexity structure that can be exploited to relax the number of variables to $2$ via Jensen's inequality. Though the relaxed program is still non-convex, we can then perform a parametric search over the two variables to find the global optimum. This allows us to compute and reasonably tightly bound the distortion for all small $k \ge 2$ in \cref{fig1}.

\subsection{Related Work}
Our model relates to three strands of recent research and we compare with them below.

\paragraph{Metric Distortion of Voting.} Building on the work of~\cite{ProcacciaR06}, the work of~\cite{AnshelevichBEPS18} initiated the study of metric distortion. Given a social choice rule with ordinal voter preferences that are consistent with an underlying metric over voters and candidates, the distortion is the worst-case (over all metrics) ratio of the total distance to the chosen alternative to that for the $1$-median alternative had the metric been known. They showed that the Copeland rule has distortion $5$, and no deterministic voting rule has distortion better than $3$.  The upper bound was subsequently improved via weighted tournament rules to $2 + \sqrt{5}$ by~\cite{MunagalaW19,Kempe_2020}, and later to $3$ by~\cite{Gkatzelis0020,Kizilkaya022}.  For randomized voting rules, the work of~\cite{AnshelevichP17} showed a lower bound of $2$. This lower bound was subsequently improved to $2.11$ by~\cite{CharikarR22}. An upper bound of $3$ follows from random dictatorship~\cite{AnshelevichP17}, and this was improved via novel voting rules to $2.74$ in~\cite{CharikarWRW24}. We refer the reader to~\cite{AnshelevichFSV21} for a survey.

Our work shows that simple models of deliberation in very small groups ($k=3,4$), where the output of the group is ordinal, yet based on their collective cardinal preferences, suffices to allow the Copeland rule to break the metric distortion lower bounds, and the distortion approaches $1$ as the group size increases. In the absence of deliberation, ``optimal'' rules such as Plurality Veto~\cite{Kizilkaya022} have better distortion than Copeland. However, these rules always output the favorite candidate of some voter. If such a rule is used to aggregate the outcome of deliberations, the distortion will be $3$ for any constant group size (see \cref{eg1}). This showcases a nice property of the Copeland rule, along with its relative ease of analysis.

\paragraph{Voting with Cardinal Information.} Our work shows that if groups of voters output ordinal preferences over alternatives via aggregating their cardinal metric information using deliberation, the resulting ordinal information can be aggregated in a way that achieves distortion bounds arbitrarily close to $1$. In our models, the outcome of deliberation favors voters with large bias towards one outcome, which is also where the median in the metric space between these voters will likely lie. 

The work of~\cite{Strength} is similar in spirit in that it tags voters' preferences between pairs of alternatives with ``strong'' and ``weak'', counting each strong preference by a fixed larger amount than a weak preference.  However, in their model, the threshold between strong and weak preference is an arbitrary threshold on the ratio between distances to the two alternatives, and the score bump is also an arbitrary number. Even with an optimal setting of these thresholds, they are unable to break the metric distortion lower bound of $3$. This is because the voting rule only gets the original ordinal information had all voters had strong or weak preferences.  Our model has two advantages. First, it does not have arbitrary thresholds. Second, in our model, voters within a group use cardinal information optimally via deliberation  to find an ordinal ranking between two outcomes (say using $1$-median). This is not only more natural, but also squeezes more information about the metric space from the voters, so that the distortion asymptotically approaches $1$.  We also refer the reader to~\cite{Distributed} for a related distributed model for small groups. . 

\paragraph{Sampling and Sortition.} Several works~\cite{FainGMS17,FainGMP19,FainFM20,CaragiannisM024} have considered distortion of voting rules when voters are randomly sampled from the population. The random dictatorship mechanism samples one voter, and achieves distortion $3$. The work of~\cite{FainGMS17} shows that if deliberation between two voters is modeled as bargaining with a disagreement alternative, then there is a protocol that achieves low distortion on special types of metrics called {\em median spaces}. The work of~\cite{FainGMP19} extends this model to three sampled voters, and bounds the second moment of distortion (and not just the expectation), which for random dictatorship, is unbounded. See also~\cite{GoelLee} for a different protocol and analysis for three voters. The work of~\cite{FainFM20} shows that sampling $k$ voters leads to bounded the $k^{th}$ moment of the distortion. However, with the exception of~\cite{FainGMS17}, these works assume voters reason with their ordinal preferences, and none of these works improve on the distortion bound of $3$. In contrast, we model deliberation in a general metric space as a natural process of reasoning with cardinal information, which leads better distortion bounds with equally small group sizes.

Motivated by citizen assemblies and sortition~\cite{Ingham18,fishkin1991democracy}, the work of~\cite{CaragiannisM024} considers a model where a large random sample of voters is chosen to deliberate and find a socially optimal outcome, or $1$-median. 
Their work makes a great case for sortition: for example, most citizen assemblies have the number of participants in the several hundreds~\cite{warren2008designing, dryzek2011toward}, and the bounds of $O((\log m)/\eps^2)$ on group size provided by~\cite{CaragiannisM024} are quite salient in this setting. However, as discussed before, research in psychology and organizational science shows that such large group sizes lead to sub-optimal deliberation. This is a substantial difficulty in applying the results of~\cite{CaragiannisM024} to design an actual deliberation process. As \cref{eg1} shows, the distortion remains lower bounded by $3$ unless the group size becomes logarithmic in the number of outcomes, so this difficulty is not just a matter of improving their analysis. Our distortion bound of $1 + O(1/k)$ becomes more salient in this setting since it allows us to use groups of size $O(1/\eps)$ to get distortion bounds of $1+\eps$. Given the impossibility result mentioned above, the process has to change in some way to obtain our bound; hence, breaking down the deliberation process so that each participant takes part in multiple small deliberations as opposed to one large one is key to our result. Thus, our result is not merely a technical improvement over~\cite{CaragiannisM024}, but is qualitatively different. In fact, many existing online platforms for synchronous deliberation (e.g.~\cite{stanfordd, myjunto, mismatch}) divide the participants into much smaller groups of 3 to 15, a range where our results are salient. Additionally, our results can be composed with those of~\cite{CaragiannisM024} to simultaneously provide a bound of $O((\log m)/\eps^2)$ on the total number of participants and of $O(1/\eps)$ on the size of each group in a single deliberation. 

Similarly, the work of~\cite{FlaniganPW23} posits a deliberation model where voters incorporate social utility into their rankings. We note that this work also does not posit an interaction mechanism in a small group. Additionally, it focuses on showing that the distortion under social welfare approaches a constant or linear for natural rules ({\em i.e.}, the social welfare approach starts to achieve bounds similar to those known for metric distortion), whereas we show improved metric distortion bounds. Thus, our results are both different from and complementary to this work as well as that of~\cite{CaragiannisM024}.

Our main contribution is thus an analytical justification for such a model of interaction with very small groups. We note that the overall number of sampled voters needed for distortion close to $1$ in the averaging model remains logarithmic in the number of alternatives, hence being comparable to the bound for a single sortition. Our approach therefore presents an alternate view of how sortition can be implemented via multiple small groups, with smoothly improving distortion bounds in the group size.

%% file: model.tex
\section{Model and Preliminaries}
\label{sec:prelim}
We now formally present the model and framework for small group interaction. There is a set $C$ of $m$ candidates (or alternatives or outcomes), and $m!$ possible rankings of these alternatives. We assume a continuum of voters, each of whose preference follows one of the rankings.\footnote{We make the continuum assumption on voters for analytic convenience, noting that the relevant lower bounds for the finite voter metric distortion problem hold in the continuum voter setting as well. (See \cref{thm:lb_main}.)} The preferences of a fraction $\rho_{q}$ of the voters follow ranking $q$. 

In the metric distortion framework, we assume the rankings with $\rho_q > 0$ and the candidates are embedded in a latent metric space with distance function $d(\cdot, \cdot)$. We assume the metric space has discrete (and finite number of) locations, corresponding to the candidates and the rankings with $\rho_q > 0$. The voters whose preferences follow ranking $q$ are placed at a set of locations $S_q$, where the voter mass at location $i \in S_q$ is $\rho_{iq}$, with $\sum_{i \in S_q} \rho_{iq} = \rho_q$.  We use ``metric space'' to mean both the distance function, as well as the voter mass at each location. We denote the combination $(d, \vec{\rho})$ as $\sigma$. Let $Q$ denote the set of all locations of voters. Given a location $i \in Q$, the ranking corresponding to $i$ is denoted $\sigma_i$, and the voter mass there is denoted $\rho_i$.

\paragraph{Classic Distortion.} In the absence of deliberation, we assume the metric space is ``consistent'' with the rankings. By this, we mean that given location $i$ with $\rho_i > 0$, and two candidates $c_1$ and $c_2$,  the ranking $q = \sigma_i$ places $c_1$ higher than $c_2$  if $d(i,c_1) < d(i,c_2)$. If $d(i,c_1) = d(i,c_2)$, then either of the two candidates can be ranked higher than the other.  Note that a consistent metric always exists by placing all rankings with $\rho_q > 0$ at location $a$ and all candidates at location $b$.

A social choice rule $\mathcal{S}$ takes the rankings with $\rho_q > 0$ (and the corresponding $\rho_q$) as input, and outputs a single candidate as the ``winner''.   Note that the social choice rule only sees the rankings and the $\rho_q$, but not the underlying metric space. 
We also assume the social choice rule is {\em anonymous} meaning that its output does not depend on the identities of the alternatives -- if the alternatives are permuted by $\pi$, the social choice rule outputs $\pi(c)$ if it was originally outputting $c$. 

To measure the quality of the social choice rule, for any candidate $c$, let $SC(c,\sigma) = \sum_{i \in Q} \rho_i d(i,c)$ denote the social cost of $c$ under the metric $\sigma= (d, \vec{\rho})$.   Let $c^*(\sigma) = \mbox{argmin}_{c \in C} SC(c,\sigma)$ be the {\em social optimum} or the median outcome had the metric space $\sigma$ been known. This outcome need not be unique, in which case an arbitrary such outcome is output.  Let $\alg(\mathcal{S}, \vec{\rho})$ denote the output of the social choice rule $\mathcal{S}$ given the rankings $q$ and their masses $\rho_q$. Then the distortion of the rule is 
$$\mbox{Distortion } = \max_{\sigma } \frac{SC(\alg(\mathcal{S}, \vec{\rho}),\sigma)}{SC(c^*(\sigma),\sigma)},$$ 
The goal is to find a social choice rule with small (preferably constant) distortion.

\subsection{Deliberation Models} 
In our deliberative framework, there is a group size parameter $k \ge 2$. A single deliberation is a function that operates over two alternatives $c_1$ and $c_2$, and $k$ randomly chosen voters from the continuum of voters, and outputs one of the alternatives. By ``randomly chosen'', we mean $k$ rankings are chosen independently with replacement from the set of rankings, with ranking $q$ being chosen at each step with probability $\rho_q$. Given the continuum assumption on voters, these $k$ rankings (some of which are possibly repeated) correspond to $k$ distinct voters.

In order to define the deliberation function, we first need to define normalized bias. Given a voter at location $i \in Q$, define the {\em normalized bias} as
$$ \B_i(c_1,c_2) = \frac{d(i,c_1) - d(i,c_2)}{d(c_1,c_2)}.$$
The numerator captures the extent of the bias towards one of the two alternatives, with a positive number denoting bias towards $c_2$. The denominator normalizes this bias by the distance between the two alternatives. 

By the triangle inequality $\B(c_1,c_2) \in [-1,1]$. Note that $\B_i(c_1,c_2) = -1$ implies $d(i,c_2) = d(i,c_1) + d(c_1,c_2)$, so that given $d(i,c_1)$ and $d(c_1,c_2)$, $d(i,c_2)$ is largest possible. This captures the maximum bias of voters towards $c_1$. Similarly $\B_i(c_1,c_2) = 1$ makes $d(i,c_2)$ as small as possible, and captures strong bias towards $c_2$. Note that when $\B(c_1,c_2) = 0$, then $d(i,c_1) = d(i,c_2)$, so the voter is indifferent between the two outcomes.

The alternative that is output by the deliberation is a function of these normalized biases for the $k$ voters. We consider the following two models. In both these models, we let $S$ be the multi-set of locations of the voters who participate in the deliberation, so that $|S| = k$.

\begin{description}
\item[Averaging Model.] (\cref{sec:avg}.) 
Let $\tau=\sum_{i\in S} B_i(c_1,c_2)$.
If $\tau<0$, the outcome is $c_1$, and if $\tau>0$, the outcome is $c_2$.
When $\tau=0$, an arbitrary tie-breaking rule is used. This corresponds to voters choosing an element of
$
\arg\min_{c\in\{c_1,c_2\}} \sum_{i\in S} d(i,c).
$
Equivalently, the group may deliberate over all $m$ alternatives and output a ranking consistent with total distance, with ties broken arbitrarily. This model is equivalent to the model where the group deliberates over all $m$ alternatives, and outputs the ranking consistent with the total distance, so that $c_1$ is ahead of $c_2$ in the ranking if $ \sum_{i \in S} d(i,c_1) \le \sum_{i \in S} d(i,c_2)$. In the case of a tie, we assume an arbitrary tie-breaking rule between the alternatives.

In the analysis below, whenever we use the event
$
\sum_{i\in S} B_i(c_1,c_2)\le 0,
$
this is only a relaxation of the actual tie-breaking rule: if $c_1$ wins with probability at least $1/2$ under any tie-breaking rule, then the probability of the weak inequality above is also at least $1/2$. For lower-bound constructions involving ties, we either split zero-sum group outcomes symmetrically or perturb the constructed distribution by an arbitrarily small amount so that zero-sum groups have probability zero.

\item[Random Choice Model.] (\cref{sec:random}.) In this model, let $S_1 \subseteq S$ denote the set of voters who prefer $c_1$ to $c_2$, and let $S_2 \subseteq S$ denote those that prefer $c_2$ to $c_1$. Let $A = \sum_{i \in S_1} |\B_i(c_1,c_2)|$ denote the total normalized bias of voters preferring $c_1$, and let $B = \sum_{i \in S_2} |\B_i(c_1,c_2)|$ be that for $c_2$.\footnote{Since $S$ is a multiset, if a location $i$ is repeated $\ell$ times, its contribution to the summation is $\ell \cdot |\B_i(c_1,c_2)|$.} Then the output is $c_1$ with probability $\frac{A}{A+B}$, and $c_2$ otherwise. 
We assume the metric is perturbed slightly so that at least one $\B_i(c_1,c_2) \neq 0$; alternatively, if all $\B_i(c_1,c_2) = 0$, we assume $c_1$ or $c_2$ is chosen with probability $1/2$ each.

In \cref{sec:random}, we also generalize the model so that there is a concave, non-decreasing function $g(x)$ with $g(0) = 0$ and $g(1) = 1$. We use $A = \sum_{i \in S_1} g(|\B_i(c_1,c_2)|)$ and $B = \sum_{i \in S_2} g(|\B_i(c_1,c_2)|)$ in randomly choosing the outcome.
\end{description}

The random choice model captures the intuition that more biased voters are likely to drive the discussion in their direction, though that outcome is far from certain. This model can also be interpreted as opinion change as follows: During deliberation, suppose every voter in $S_1$ independently changes their opinion to $c_2$ with probability $\frac{B}{A+B}$, and similarly, every voter in $S_2$ changes their opinion to $c_1$ with probability $\frac{A}{A+B}$. This process corresponds to the DeGroot model of opinion change \cite{degroot1974reaching}. If subsequently, the opinion of a random voter is implemented, this exactly corresponds to the random choice model. A final, somewhat non-deliberative interpretation is that the process chooses a voter proportional to their bias (which might be proportional to how much they speak up), and outputs that voter's preference. 

\paragraph{Remark.} For $k = 1$, both models reduce to a single voter choosing the candidate among $c_1, c_2$ that is higher in their ranking. Note also that in both these models, the outcome of deliberation is simply an ordinal preference between the two alternatives. The voters arrive at this alternative by reasoning about the underlying metric space, but this reasoning is hidden from the social choice rule that we describe next.

%% file: copeland.tex
\subsection{Normalized Bias and the Distortion of Copeland's Rule}
\label{sec:copeland}
We now define the social choice rule to aggregate the outcome of deliberations. Note that each deliberation is a probabilistic mapping of a pair of alternatives to a winning alternative. This mapping depends on the underlying metric space as well as the $k$ randomly sampled voters. 

Given a deliberation model $\A$ and metric space $\sigma = (d, \vec{\rho})$, let $p_k(W,X)$ denote the probability that the outcome of deliberation among a set $S$ of $k$ randomly chosen participants between alternatives $W$ and $X$, outputs $W$ instead of $X$. This probability is over both the randomness in the choice of $S$ and the randomness in the outcome of the deliberation given $S$ (as in the random choice model). To keep our analysis simple, we assume the probability can be exactly estimated for any $(W,X)$; in other words, we assume infinitely many groups of size $k$ can be drawn from the population, and small group deliberations  run  between $W$ and $X$.  In \cref{thm:sample2,thm:sample1}, we will remove this assumption and analyze the number of groups of size $k$ that need to be sampled to approximately achieve the same distortion.

\paragraph{Copeland Rule and Distortion.} The Copeland rule dates back to Ramon Llull in the $13^{th}$ century~\cite{voting-book}. To define this rule, we say that $W \succ X$ if $p_k(W,X) \ge 1/2$.  
For every pair $(X,Y)$, the Copeland rule gives $1$ point to $X$ (resp. $Y$) if $p_k(X,Y) > 1/2$ (resp. $p_k(Y,X) > 1/2$), and half a point to each alternative if $p_k(X,Y) = 1/2$. It then chooses the alternative $W$ with most points.  Such an alternative $W$ belongs to the {\em uncovered set}~\cite{landau}: For any other alternative $X$, either $W \succ X$, or there is an alternative $Y$ such that $W \succ Y$ and $Y \succ X$. We assume an arbitrary alternative in the uncovered set is output. Note that the Copeland rule operates over the ordinal outcomes of the deliberations, without knowing the metric space $\sigma$.

To define distortion, 
let $\alg(\A,\sigma)$ denote the outcome of applying Copeland to the tournament graph on deliberations using model $\A$. As before, let $c^*(\sigma) = \mbox{argmin}_{c \in C} SC(c,\sigma)$ denote the social optimum. Then the distortion of the deliberation rule $\A$ under the Copeland rule is:

$$\mbox{Distortion}(\A) = \max_{\sigma}  \frac{SC(\alg(\A,\sigma), \sigma)}{SC(c^*(\sigma),\sigma)}.$$

If the outcome $\alg(\A,\sigma)$ is probabilistic, we replace the numerator in the above expression by its expectation over the randomness in $\A$.

\paragraph{The Averaging Model is Deterministic.} As an aside, suppose there is a finite number $n$ of voters, where $n$ is suitably large. Then, in the Averaging model, the randomness in the choice of set $S$ can be replaced by considering {\em all} subsets $S$ of $k$ voters, and computing $r_k(W,X)$ as the fraction of these sets where deliberation between alternatives $W$ and $X$ leads to $W$ being chosen.  It is easy to check that as $n \rightarrow \infty$, the value $r_k(W,X)$ converges to $p_k(W,X)$. Note that the Averaging model itself is a deterministic process given $S$; further, the Copeland rule applied to the $\{r_k(W,X)\}$ is deterministic. We therefore compare the distortion of the Averaging model with that of {\em deterministic} social choice rules without deliberation. In the subsequent discussion, we will go back to assuming there is a continuum of voters, and $S$ is a randomly chosen set of $k$ rankings.

\paragraph{Mathematical Program for Distortion.} We now show a simple mathematical program whose optimal solution yields an upper bound on distortion of the Copeland rule.

Fix a metric $(d ,\vec{\rho})$, and consider two alternatives $W$ and $X$. Let $\phi_i = \B_i(W,X)$ be the normalized bias of voters at location $i$, and let $D$ denote the distribution of $\phi$, so that $\phi = \phi_i$ with probability $\rho_i$. Since we fix the metric space, we omit it from the notation $SC(W)$, etc. We have the following lemma, which formulates the analysis of the Copeland rule in~\cite{AnshelevichBEPS18} as the solution to a mathematical program.

\begin{lemma} Let $\gamma = \E[D]$. If $\gamma\le \theta$ for some $\theta\in[0,1)$, then
\[
\frac{SC(W)}{SC(X)}\le \frac{1+\theta}{1-\theta}.
\]
\end{lemma}
\begin{proof}
Assume $d(W,X) = 1$. First assume $\gamma \ge 0$. Then we have 
$$SC(W) - SC(X) = \sum_i \rho_i \left(d(i,W) - d(i,X) \right) = \sum_i \rho_i \phi_i = \E[D] = \gamma. $$
Next, by triangle inequality, $d(i,X) \ge d(X,W) - d(i,W)$, so that
$$ SC(X) = \sum_i \rho_i d(i,X) \ge \frac{1}{2} \sum_i \rho_i \left(d(X,W) - (d(i,W) - d(i,X) \right) = \frac{1}{2} \left(1 - \gamma \right). $$
Therefore, 
\[
\frac{SC(W)}{SC(X)}
=1+\frac{\gamma}{SC(X)}
\le 1+\frac{2\gamma}{1-\gamma}
=\frac{1+\gamma}{1-\gamma}.
\]

If $\gamma<0$, then $SC(W)=SC(X)+\gamma<SC(X)$, so $SC(W)\le SC(X)$.
\end{proof}

We now find the worst case of $\gamma$ over all metrics $\sigma = (d, \vec{\rho})$. For this, let $q(W,X)$ denote the solution to the following optimization problem: Find a distribution $D = \{\rho_i\}$ over $\{\B_i(W,X)\}$ which optimizes:
\begin{equation} 
\label{eq:opt}
\max_D \E[D] \qquad \mbox{s.t.} \qquad W \succ X,
\end{equation}
Since $\B_i(W,X) \in [-1,1]$, the above can be reformulated as optimizing over all distributions $D$ supported on $[-1,1]$.  We next show that upper-bounding this optimum  suffices to upper-bound the distortion. 

\begin{theorem}
\label{thm:distort1}
For deliberation model $\A$, if $q(W,X) \le \theta$ for all pairs $W,X$ of alternatives, then the distortion of the Copeland rule applied on the outcomes of $\A$ is at most $ \left( \frac{1 + \theta}{1-\theta}\right)^2$.
\end{theorem}
\begin{proof}
    Fix some metric $(d,\vec{\rho})$. Suppose the Copeland rule finds outcome $W$, and suppose $X$ is the social optimum. In the case where $W \succ X$, by the previous lemma, the distortion is 
    $$\frac{SC(W)}{SC(X)} \le \frac{1 + \theta}{1-\theta}.$$ 
    Otherwise, there exists $Y$ such that $W \succ Y$ and $Y \succ X$. Then, the distortion is 
    $$ \frac{SC(W)}{SC(X)} = \frac{SC(W)}{SC(Y)} \cdot \frac{SC(Y)}{SC(X)} \le  \left( \frac{1 + \theta}{1-\theta}\right)^2. $$
    Since these bounds hold for all metrics $(d,\vec{\rho})$, the proof is complete.
\end{proof}

Given the above theorem, for any deliberation model $\A$,  our goal is to upper bound the optimum $q(W,X)$ in \cref{eq:opt} for any pair of alternatives. This will imply an upper bound on distortion. Since this upper bound is independent of the alternatives themselves (as it simply optimizes over all distributions supported on $[-1,1]$), we only need to show it for an arbitrary pair of outcomes. We will do this in the next sections for the averaging and the random choice deliberation models.

\paragraph{A Lower Bound.}
Before proceeding further, we note that solving \cref{eq:opt} also provides a lower bound for distortion. This lower bound uses only two candidates and applies to any social choice rule that only aggregates the ordinal rankings output by all deliberating groups of size $k$.

We will assume only two candidates $W,X$ in the proof below. For a pairwise deliberation model, let $p(D)$ denote the probability that a group of size $k$ ranks $W$ above $X$ when the normalized biases $B_i(W,X)$ of the group members are drawn i.i.d. from the distribution $D$ over $[-1,1]$, where positive bias favors $X$ and negative bias favors $W$. 

In the two candidate setting, we make the following mild assumption about the deliberation model, which can be easily made to hold for the deliberation models considered in this paper: 

\begin{quote}
    {\bf Reversal property.} For every $\mu<\theta_k$, there is a distribution $D$ over $[-1,1]$ such that $\E[D]\ge \mu$ and $p(D)=p(-D)=1/2$.
\end{quote} 
For completeness, we justify this property for our models. Start with a feasible distribution $D_0$ for \cref{eq:opt} such that $\E[D_0] > \mu$. In the averaging model, we suppose the mass of ties (where the $k$-tuple equally prefers $W$ to $X$) is equally split between $W \succ X$ and $X \succ W$. For the random choice model, we smoothen $D_0$ slightly so that the all-zero group has measure zero. This changes the expectation and $p(\cdot)$ by an arbitrarily small amount, so the expectation remains larger than $\mu$ and feasibility is preserved. If now $p(D_0)=1/2$, then reversal of biases gives $p(-D_0)=1/2$, and we are done. Otherwise $p(D_0)>1/2$. Consider $D_t=(1-t)D_0+t\delta_1$ for $t\in[0,1]$, where $\delta_1$ is the point mass at $1$. This continuously moves probability mass toward the largest positive bias. It weakly increases the expectation and weakly decreases $p(\cdot)$, since larger biases favor $X$. Also $p(\delta_1)<1/2$. Therefore, by continuity, there is some $t^\star$ such that $p(D_{t^\star})=1/2$, while still $\E[D_{t^\star}]\ge\mu$.  Rename $D_{t^\star}$ as $D$ and we are done.

\begin{theorem}
\label{thm:lb_main}
Fix $k \ge 2$. Suppose the social choice rule satisfies the reversal property above, and given only the distribution of ordinal rankings output by all groups of size $k$, aggregates these rankings to choose an alternative. If \cref{eq:opt} has value $\theta_k$, then the distortion of any deterministic social choice rule is at least $\min\left(3,\frac{1+\theta_k}{1-\theta_k}\right)$, and that of any randomized social choice rule is at least $\min\left(2,\frac{1}{1-\theta_k}\right)$.
\end{theorem}

\begin{proof}
If $\theta_k\le 0$, the claimed lower bounds are trivial, so assume $\theta_k>0$. Fix any $0<\mu<\min\{\theta_k,1/2\}$. Using the reversal property, choose a distribution $D$ over normalized biases such that $\E[D]\ge\mu$ and $p(D)=p(-D)=1/2$.

We construct two metric instances with two alternatives $W$ and $X$ at distance $1$ from each other on a line. In the first instance, denoted $\sigma^+$, suppose $\Pr[D=a]=p_a$ for $a\in[-1,1]$. Place voters of mass $p_a$ at distance $\frac{1+a}{2}$ from $W$ and distance $\frac{1-a}{2}$ from $X$. The normalized bias of these voters for the ordered pair $(W,X)$ is exactly $a$. Hence the distribution of group-output rankings has a $1/2$ fraction ranking $W$ above $X$ and a $1/2$ fraction ranking $X$ above $W$. Writing $\bar\mu=\E[D]\ge\mu$, we have $SC_{\sigma^+}(W)=\frac{1+\bar\mu}{2}$ and $SC_{\sigma^+}(X)=\frac{1-\bar\mu}{2}$. Thus $X$ is socially optimal in $\sigma^+$, and choosing $W$ incurs distortion
$$
\frac{SC_{\sigma^+}(W)}{SC_{\sigma^+}(X)}
=
\frac{1+\bar\mu}{1-\bar\mu}
\ge
\frac{1+\mu}{1-\mu}.
$$

In the second instance, denoted $\sigma^-$, use the reflected distribution $-D$: voters of mass $p_a$ are placed at distance $\frac{1-a}{2}$ from $W$ and distance $\frac{1+a}{2}$ from $X$. Since $p(-D)=1/2$, this instance induces exactly the same distribution of ordinal rankings output by all groups of size $k$ as $\sigma^+$. However, the social costs are reversed: $SC_{\sigma^-}(W)=\frac{1-\bar\mu}{2}$ and $SC_{\sigma^-}(X)=\frac{1+\bar\mu}{2}$. Thus $W$ is socially optimal in $\sigma^-$, and choosing $X$ incurs distortion at least $\frac{1+\mu}{1-\mu}$.

First consider deterministic rules. Since $\sigma^+$ and $\sigma^-$ induce the same input to the social choice rule, any deterministic rule must output the same alternative on both instances. If it outputs $W$, then $\sigma^+$ gives distortion at least $\frac{1+\mu}{1-\mu}$. If it outputs $X$, then $\sigma^-$ gives distortion at least $\frac{1+\mu}{1-\mu}$. Therefore every deterministic rule has distortion at least $\frac{1+\mu}{1-\mu}$. Since $\mu<\min\{\theta_k,1/2\}$ was arbitrary, the deterministic lower bound is $\min\left(3,\frac{1+\theta_k}{1-\theta_k}\right)$.

For randomized rules, let $\alpha$ denote the probability that the rule outputs $W$ on this common input, and let $R=\frac{1+\bar\mu}{1-\bar\mu}$. On $\sigma^+$, the expected distortion is $\alpha R+(1-\alpha)$, while on $\sigma^-$, it is $\alpha+(1-\alpha)R$. The larger of these two quantities is at least their average:
$$
\frac{1}{2}\left(\alpha R+(1-\alpha)+\alpha+(1-\alpha)R\right)
=
\frac{1+R}{2}
=
\frac{1}{1-\bar\mu}
\ge
\frac{1}{1-\mu}.
$$
Since $\mu<\min\{\theta_k,1/2\}$ was arbitrary, the randomized lower bound is $\min\left(2,\frac{1}{1-\theta_k}\right)$.
\end{proof}

%% file: average.tex
\section{Distortion in the Averaging Model}
\label{sec:avg}
We first consider the averaging model. We show in \cref{thm:berry} that the group size of $k = \Theta(1/\epsilon)$ is both necessary and sufficient for distortion $1+\epsilon$. In this model, it is analytically difficult to tightly bound distortion for small $k$; nevertheless, we show that the distortion is tight at $3+\sqrt{2} \approx 4.414$ for groups of size $k=2$ (\cref{thm:theta2}), while it is at most $2.78$ for groups of size $k=3$ (\cref{thm:theta3}). 
This shows that a very small group size ($k=3$) suffices to beat the deterministic metric distortion lower bound of $3$ in this model. (Note that by the discussion in \cref{sec:prelim}, the Averaging model can be viewed as a deterministic social choice rule.) We finally show in \cref{thm:sample2} that the number of sampled groups needed to achieve an additive $\delta$ approximation to this bound is only $\tilde{O}(\log m)$, where $m$ is the number of alternatives. 

In the averaging model, we choose a set of $k$ random voters to deliberate between two alternatives $W$ and $X$. Let the multiset of their locations be denoted by $S$. We compute $\sum_{i \in S} \B_i(W,X)$. If this is negative, the outcome of the deliberation is $W$, else it is $X$. This corresponds to outputting $\mbox{argmin}_{c \in \{W,X\}} \sum_{i \in S} d(i,c)$.  This model is equivalent to the model where the group deliberates over all $m$ alternatives, and outputs the ranking consistent with the total distance. This implies $c_1$ is ahead of $c_2$ in the ranking if $ \sum_{i \in S} d(i,c_1) \le \sum_{i \in S} d(i,c_2)$, and our analysis extends directly to that case.

We now specialize \cref{eq:opt} for this model. Suppose there are two alternatives $W,X$ between which the deliberation happens. Let $\phi_i = \B_i(W,X)$. The event of $W$ being the winner maps to the condition $\sum_{i \in S} \phi_i \le 0$. Therefore, we seek a distribution $D$ whose support is $[-1,1]$ that solves the following problem:
\begin{equation}
\label{eq:opt_avg}
\theta_k :=  \ \ \max_{D} \E[D] \qquad \mbox{s.t.} \qquad \Pr_{\{a_1,a_2,\ldots,a_k\} \sim D} \left[ \sum_{i=1}^k a_i \le 0 \right] \ge 1/2. 
\end{equation}
where the probability is over a set of $k$ independent samples $a_1,\ldots,a_k$ drawn from $D$.

\subsection{Asymptotic Behavior of Distortion in Group Size $k$}
\label{sec:berry}
We will now use the Berry-Esseen theorem~\cite{feller1971introduction} to show a tight bound of  on the asymptotic behavior of $\theta_k$ in $k$. This will yield the following theorem.

\begin{theorem}
\label{thm:berry}
For any $k \ge 2$,  $\theta_k = O\left(\frac{1}{k}\right)$. By \cref{thm:distort1}, this means the distortion of the Copeland rule in the averaging model of deliberation using groups of size $k$ is $1 + O\left(\frac{1}{k}\right)$. In particular, for any $\epsilon > 0$, groups of size $k = O\left(\frac{1}{\epsilon}\right)$ are sufficient for distortion $1+\epsilon$.  
\end{theorem}
\begin{proof}
Let $\theta_k = \mu = \E[D] \ge 0$, and let $D_k = \frac{\sum_{i=1}^k a_i}{k}$, so that $\E[D_k] = \mu$. Let $\sigma^2 = \mbox{Var}[D]$.  If $\sigma=0$, then $D$ is almost surely equal to $\mu$. Since $\Pr[D_k\le 0]\ge 1/2$, we must have $\mu\le 0$. But the point mass at $0$ is feasible, so in the optimizing case with $\mu=\theta_k\ge 0$, this gives $\mu=0$, and the claim is immediate. Hence assume $\sigma>0$.

We have $\mbox{Var}[D_k] = \sigma^2/k$. The condition $W \succ X$ implies $\Pr[D_k \le 0] \ge 1/2$, which means the median is at most $0$. Noting that the gap between the median and mean of any distribution is at most the standard deviation~\cite{MU}, this implies 
$$\mbox{Var}[D_k] = \frac{\sigma^2}{k} \ge (0 - \mu)^2  \qquad \Rightarrow \qquad \sigma \ge \mu \sqrt{k}.$$
Since $\sigma \le 1$ for a random variable bounded in $[-1,1]$, this already implies $\theta_k = \mu \le \frac{1}{\sqrt{k}}$. 

We will now improve this bound using the Berry-Esseen inequality~\cite{feller1971introduction} applied to the average of $k$ random variables distributed as $D - \mu$. Note that $\E[D-\mu] = 0$, and $\E[(D-\mu)^2] = \mbox{Var}[D]= \sigma^2$. Further since $\mu, |D| \le 1$, we have $\E[|D|^3] \le \E[D^2]$, so that
 \begin{align*}
    \E[|D-\mu|^3]  \le & \ \E[(|D| + \mu)^3] \\
     = &\  \E[|D|^3] + 3 \mu \cdot \E[D^2] + 3 \mu^2 \cdot \E[|D|] + \mu^3 \\
     \le &\  4 \cdot \E[D^2] + 4 \cdot \mu^2 \\
    = & \ 4 \cdot \sigma^2 + 8  \cdot \mu^2 
\end{align*} 

Let $Y = \frac{(D_k - \mu) \sqrt{k}}{\sigma}$, so that $\E[Y] = 0$ and $\mbox{Var}[Y] = 1$. The condition $W \succ X$ implies 
$$\Pr\left[Y \le \frac{-\mu \sqrt{k}}{\sigma} \right] \ge 1/2.$$

Let $F_Y(z)$ denote $\Pr[Y \le z]$, and $\Phi(z)$ denote the CDF of the standard Normal distribution.  Let $z^* = -\frac{\mu \sqrt{k}}{\sigma}$.  By the Berry-Esseen theorem  (whose preconditions can be checked to apply here) applied at the point $z^*$, we have:
$$   \frac{1}{2} - \Phi(z^*) \le F_Y(z^*) - \Phi(z^*)  \le \frac{0.5 \cdot E[|D - \mu|^3]}{\E[(D-\mu)^2]^{3/2} \cdot \sqrt{k}} \le  \frac{2 \sigma^2 + 4 \mu^2}{\sigma^3 \sqrt{k}}.$$
where we have used $\E[|D-\mu|^3] \le 4 \sigma^2 + 8 \mu^2$.

Next note that by the above lower bound on $\sigma$, we have $z^* \ge -1$, so that $\phi(z^*) \ge \frac{1}{\sqrt{2\pi e}}$, where $\phi$ is the PDF of the standard Normal distribution. We therefore have
$$   \frac{1}{2} - \Phi(z^*) = \Phi(0) - \Phi(z^*)  \ge \phi(z^*) \cdot (0 - z^*) \ge \frac{1}{\sqrt{2\pi e}} \cdot \frac{\mu \sqrt{k}}{\sigma}.$$
Putting the above two inequalities together, and using $\frac{\mu^2}{\sigma^2} \le \frac{1}{k}$, we have
$$ \frac{1}{\sqrt{2\pi e}} \cdot\frac{\mu \sqrt{k}}{\sigma} \le \frac{2 \sigma^2 + 4 \mu^2}{\sigma^3 \sqrt{k}} \qquad \Rightarrow \qquad \mu \le 2  \sqrt{2\pi e} \cdot \left(\frac{1}{k} + \frac{2}{k^2}  \right) \le \frac{8.27}{k} \cdot \left(1 + \frac{2}{k}\right).$$
Since $\theta_k = \mu$, plugging this bound into \cref{thm:distort1}, the proof of the upper bound is complete.
\end{proof}

We now show a matching lower bound, showing our analysis is tight.

\begin{theorem}
\label{thm:lb1}
For any $k \ge 2$, $\theta_k = \Omega\left(\frac{1}{k}\right)$. This means that any anonymous social choice rule that aggregates the ordinal rankings of all groups of size $k$ over all alternatives has distortion $1 + \Omega\left(\frac{1}{k}\right)$, so that achieving distortion $1+\epsilon$ needs groups of size $\Omega(1/\epsilon)$.
\end{theorem}
\begin{proof} 
We show the lower bound on $\theta_k$. The bound on distortion follows from \cref{thm:lb_main}. The result for $k = 2$ follows from \cref{lem:theta2}.

\paragraph{Case 1. Odd $k$.} We place a mass of $1/2$ voters at $X$, and $1/2$ mass of voters between $X$ and $W$, at a distance $\frac{1}{k+1}$ from $W$. The distribution $D$ is therefore $1$ with probability $1/2$ and $-1+\frac{2}{k+1}$ with probability $1/2$, so that $\E[D] = \frac{1}{k+1}$. To see that $W \succ X$, if we sample $k$ times from this distribution, the sum of the values is at most $0$ as long as the majority of samples are closer to $W$, which happens with probability exactly $1/2$. 

\paragraph{Case 2. Even $k \ge 4$.} We place a voter mass of $p = \frac{1}{2} + \frac{1}{3k}$ at $X$ (location $1$ in distribution $D$) and $1-p = \frac{1}{2} - \frac{1}{3k}$ at $W$ (location $-1$ in  $D$). This implies $\E[D] =  \frac{2}{3k}$. To see that $W \succ X$, let $M$ denote the median of {\tt Binomial}$(k,p)$, whose mean is $\mu = k p$. The work of~\cite{Binomial} upper bounds the gap between the mean and median for a Binomial and implies that 
$$ | M - kp| = \left|M - \frac{k}{2} - \frac{1}{3}\right|  \le \max(p,1-p) \le \frac{1}{2} + \frac{1}{3k} \le \frac{7}{12} \qquad \Rightarrow \qquad \left|M - \frac{k}{2} \right| < 1.$$
Since $k$ is even, this implies $ M = \frac{k}{2}$. Consider a sample of $k$ voters. If at least $k/2$ voters are located at $W$, then $W$ is the winner. But the probability of this event is at least the probability that {\tt Binomial}$(k,p) \le M$, which is at least $1/2$. Therefore, $W \succ X$. 

The constructions above use weak inequalities at one boundary point. If the tie-breaking convention does not select $W$ at equality, perturb the construction by an arbitrarily small amount: in the odd case, move the negative atom slightly downward, and in the even case, move the positive atom from $1$ to $1-\varepsilon$. For sufficiently small $\varepsilon=o(1/k)$, the relevant sums become strictly negative, and the expectation changes only by $o(1/k)$. Thus the same $\Omega(1/k)$ lower bound holds under any fixed
tie-breaking rule, completing the proof of the lower bound. 
\end{proof}


In \cref{thm:berry}, we note that moment based methods like Bernstein's inequality~\cite{Boucheron} would only show an upper bound of $\theta_k = O\left(\frac{1}{\sqrt{k}}\right)$ -- these methods require the variance to be small, while the Berry-Esseen inequality shows the distribution of the average is closer to Normal if the variance is high while the third moment is bounded. Though the latter yields a much stronger bound, it is still far from tight for small $k$. For instance, it implies that we need a very large constant $k$ to achieve distortion below $3$, while we show below that the correct bound is $k = 3$. We next show vastly improved distortion bounds for the canonical cases of $k = 2,3$. For $k=3$, we obtain $2.78$, which beats the deterministic metric distortion lower bound. 

%% file: average_two.tex
\subsection{Distortion Bound for $k=2$}
\label{sec:two}
We now consider the case with groups of size $k=2$. We note that \cref{thm:lb_main} combined with \cref{lem:theta2} below shows a lower bound of $1+\sqrt{2}$ on the distortion of any deterministic social choice rule that uses the deliberation outcomes of all pairs of voters. We now show that Copeland has distortion $3+\sqrt{2}$, and this bound is tight. Though the $k=2$ case  sounds simple, even here, we need to solve an interesting non-convex program to global optimality. Note that our bound significantly improves the tight distortion bound of $5$ for Copeland when $k = 1$ (no deliberation).

\begin{theorem}
\label{thm:theta2}
For the Copeland rule applied to the averaging model with group size $k = 2$, the distortion is at most $3 + \sqrt{2} \le 4.42$, and this bound is tight for  rules that choose from the uncovered set.
\end{theorem}

The rest of this subsection is devoted to proving the above theorem. 

\paragraph{Computing $\theta_2$.} We first compute $\theta_2$ explicitly. We show the following lemma using the idea of pipage rounding~\cite{Ageev2004}.

\begin{lemma}
\label{lem:pipage}
    Consider \cref{eq:opt_avg} with $k = 2$. The optimal $D$ has support $\{-1,0,1\}$.
\end{lemma}
\begin{proof}
We prove the claim for distributions supported on $ G_\varepsilon=\{0,\pm \varepsilon,\pm 2\varepsilon,\ldots,\pm 1\},$
and then let $\varepsilon\to 0$.

Fix such a distribution $D$.  For an interior grid value $a\in(0,1)$, write $p=\Pr[D=a]$ and $q=\Pr[D=-a].$
Assume $p+q>0$.  We perform the following rounding operation.

If $p>q$, move the mass at $a$ to $b$ and the mass at $-a$ to $-b$, where $b>a$ is the smallest grid value such that either $b=1$ or at least one of $b,-b$ already has positive mass.  If $p\le q$, move the mass at $a$ to $b$ and the mass at $-a$ to $-b$, where $b<a$ is the largest grid value such that either $b=0$ or at least one of $b,-b$ already has positive mass; when $b=0$,
both masses are moved to $0$.

We claim that this operation does not decrease $\Pr_{x,y\sim D}[x+y\le 0]. $ Indeed, no occupied grid point has absolute value strictly between $a$ and $b$. Thus, when an atom is moved, the truth value of $x+z\le 0$ for any other occupied grid point $z$ cannot change from true to false.  The only possible new boundary cases occur when the moved atom meets an already occupied opposite atom, and these only turn a strict inequality into equality.  Since the constraint uses the weak inequality $x+y\le 0$, this cannot decrease the left-hand side.  The contribution of pairs involving the two moved atoms with
each other also remains at equality.

The objective weakly increases.  If $p>q$, then $b>a$, and the change in expectation is
$
(p-q)(b-a)>0.
$
If $p\le q$, then $b<a$, and the change in expectation is
$
(p-q)(b-a)\ge 0.
$
Therefore the operation preserves feasibility and weakly increases the objective.

Each operation removes the interior absolute value $a$ from the support and
moves its mass either to $0$, to $\pm 1$, or to an already occupied absolute
value.  Hence the number of occupied interior absolute values decreases.  After
finitely many operations, all mass is supported on $\{-1,0,1\}$, feasibility is
preserved, and the expectation has not decreased.  Letting
$\varepsilon\to 0$ completes the proof.
\end{proof}

We can now write \cref{eq:opt_avg} as a non-convex optimization problem. Suppose $\Pr[D = -1] = p$, and $\Pr[D = 0] = q$. Then the program becomes:
\begin{equation}
\max_{p,q \ge 0, p+q \le 1} 1 - 2p - q \qquad \mbox{s.t.} \qquad (p+q)^2 + 2 p (1-p-q) \ge 1/2.
\end{equation}

We now use the global optimization tool  BARON ((Branch-And-Reduce Optimization Navigator)~\cite{Sahinidis1996,KS18})~\cite{Sahinidis1996}\footnote{BARON can be called from Python using the AMPL API ({\tt amplpy}).  We have used the demo (free) version available on Google Colab, that supports $10$ constraints and variables and up to $50$ nonlinear operations. This suffices for all our programs. }.  BARON uses a combination of convex relaxation and integer programming techniques to find increasingly tighter lower and upper bounds on the global optimum. We find that $q = 0$ and $p = 1 - 1/\sqrt{2}$. Therefore,

\begin{lemma}
\label{lem:theta2}
    $\theta_2 = \sqrt{2} - 1$, and is achieved when $\Pr[D = 1] = \frac{1}{\sqrt{2}}$ and $\Pr[D = -1] = 1 - \frac{1}{\sqrt{2}}$. 
\end{lemma}

The above lemma implies that when $W \succ X$, we have $\frac{SC(W)}{SC(X)} \le \frac{1+\theta_2}{1-\theta_2} \le \sqrt{2} + 1 \approx 2.414$. It also implies a lower bound of $\sqrt{2} + 1$ on the distortion of any deterministic social choice rule by \cref{thm:lb_main}.

\paragraph{Distortion of Copeland.} We next bound $\frac{SC(W)}{SC(X)}$ when there exists a $Y$ such that $W \succ Y \succ X$. Towards this end, let $a_i = d(i,W) - d(i,Y)$ and $b_i = d(i,Y) - d(i,X)$. Let $d(Y,X) = 1$, and $d(W,X) = B \ge 0$, so that $d(W,Y) \le B+1$. By triangle inequality, we have $|a_i| \le B+1$, and $|b_i| \le 1$ for all locations $i$. 

Let $D_1$ be the distribution on $\{a_i\}$ induced by $D$, and $D_2$ be the corresponding distribution on $\{b_i\}$.  The conditions $W \succ Y$ and $Y \succ X$ yield:
\begin{equation}
\label{eq:cond1}
\Pr_{\{a_1,a_2\} \sim D_1} \left[ a_1 + a_2 \le 0 \right] \ge 1/2, \mbox{ and }
\end{equation}
\begin{equation}
\label{eq:cond2}
\Pr_{\{b_1,b_2\} \sim D_2} \left[ b_1 + b_2 \le 0 \right] \ge 1/2,
\end{equation}
where the two draws from $D_1$ (resp. $D_2$) are {\em i.i.d.} Note that $a_i + b_i = d(i,W) - d(i,X)$ so that 
\begin{equation}
\label{eq:cond_diff}
    SC(W) - SC(X) = \E[a_i + b_i].
\end{equation}
We will bound $SC(X)$ as follows. Note that
$$2 d(i,X) = d(i,X) + d(i,X) \ge (d(W,X) - d(i,W)) + d(i,X) = d(W,X) - (a_i + b_i).$$  
Similarly, we also get 
$$2 d(i,X) \ge d(Y,X) - b_i.$$
Therefore, 
\begin{equation}
\label{eq:cond_x}
    SC(X) \ge \frac{1}{2} \cdot  \E\left[\max ( B - a_i - b_i, 1 - b_i) \right].
\end{equation}
To show that the distortion is at most $1+\beta$ it therefore suffices to show
$$ SC(W) - SC(X) \le \beta \cdot SC(X).$$
Plugging in \cref{eq:cond_diff,eq:cond_x}, and simplifying, it suffices to show that the following objective is at most $0$.
\begin{equation}
\label{eq:obj4}
\eta_2 := \mbox{Maximize} \left( \E\left[\left(\frac{2}{\beta} + 1 \right) b_i \right] + \E\left[ \frac{2}{\beta} \cdot a_i +  \min\left( a_i - B + 1,  0 \right) -1\right] \right)
\end{equation}
where the maximization is over $B$, and  distributions $D_1$, $D_2$, where $D_1$ is a distribution over $\{a_i\}$ supported on $[-B-1,B+1]$, and $D_2$ is a distribution over $\{b_i\}$ supported on $[-1,1]$. The maximization is subject to constraints \cref{eq:cond1,eq:cond2}. We will now show that the above maximum is at most $0$.

Note that the optimum for $\{b_i\}$ subject to \cref{eq:cond2} and $|b_i| \le 1$ and that for $\{a_i\}$ subject to \cref{eq:cond1} and $|a_i| \le B+1$ can be separately computed. For $\{b_i\}$, the optimum is simply $(1 + 2/\beta) \cdot \theta_2$. For $\{a_i\}$, we have the following lemma.

\begin{lemma}
    The optimum distribution $D_1$ for \cref{eq:obj4} is supported on $\{-B-1, B+1, B-1, 1-B, 0\}$. 
\end{lemma}
\begin{proof}
    We use the same procedure as in the proof of \cref{lem:pipage}. For any $a > 0$, consider the objective in \cref{eq:obj4} for the points $\{a,-a\}$. Let $p = \Pr[D_1 = a]$ and $q = \Pr[D_1 = -a]$, where $\max(p,q) > 0$. The objective is a linear function of $a$ except at the point $a = B-1$, where $-a = 1-B$. To see this, note that the contribution to the objective is both linear in $a$ and linear in $-a$. As an example, if $a >  B-1$, then the contribution to the objective is 
    $$p \cdot \left( \frac{2}{\beta} \cdot a + 0 \right) + q \cdot \left( \frac{2}{\beta} \cdot (-a) +   (-a-B+1) \right), $$
    which is linear in $a$. Therefore, we can perform the same operation as in \cref{lem:pipage} and move the probability mass to the neighboring points, till either it hits one of $B-1, 1-B$, or the extreme points $\{-B-1, B+1\}$ or $0$. Therefore, the optimal $D_1$ is supported on these five points.
\end{proof}

We can now write \cref{eq:obj4} as a non-linear program with six variables, one for $B$, and the rest capturing the probabilities of $D_1$ taking one of the five support values.  Note that to encode \cref{eq:cond1}, we need to consider two cases, depending on whether $B \ge 1$ or otherwise. We elaborate below.

\paragraph{Non-convex Program: Case 1.} We first consider the case where  $0 \le B \le 1.$ Note that $B$ itself is a variable. Let 
$$p = \Pr[D_1 = B+1], q = \Pr[D_1= 1-B], r = \Pr[ D_1 = 0], s = \Pr[D_1 = B-1], t = \Pr[D_1 = -B-1].$$ 
These values of $D_1$ are in decreasing order. Then
\begin{equation} 
\label{eq:sum1}
p + q + r + s + t = 1, \qquad \mbox{and} \qquad p,q,r,s,t \ge 0.
\end{equation}
The probability constraint \cref{eq:cond1} implies
\begin{equation} 
\label{eq:prob1} (p + q)^2 + 2 p \cdot (r + s) + 2  q \cdot r \le 1/2.
\end{equation}
Let $A = \left(1 + \frac{2}{\beta}\right) \cdot \theta_2$. Then subject to the above constraints, the objective is:
$$ A +  p \left(\frac{2}{\beta}(B+1) - 1\right) + q \left(\frac{2}{\beta}(1-B) - 1\right) - r - s\left( \frac{2}{\beta} (1-B)+1\right) - t \left( \frac{2}{\beta}(B+1) + 2B + 1 \right). $$

\paragraph{Non-convex Program: Case 2.} We next consider the case where 
 $B \ge 1,$ where $B$ is a variable. Let 
$$p = \Pr[D_1 = B+1], q = \Pr[D_1= B-1], r = \Pr[ D_1 = 0], s = \Pr[D_1 = 1-B], t = \Pr[D_1 = -B-1].$$ 
These values of $D_1$ are in decreasing order. It is easy to check that constraints \cref{eq:sum1,eq:prob1} remain the same.
Let $A = \left(1 + \frac{2}{\beta}\right) \cdot \theta_2$. Then, subject to the above constraints, the objective is now:
$$ A +  p \left(\frac{2}{\beta}(B+1) - 1\right) + q \left(\frac{2}{\beta}(B-1) - 1\right) - r B - s\left( \frac{2}{\beta} (B-1)+2B-1\right) - t \left( \frac{2}{\beta}(B+1) + 2B + 1 \right). $$

Note that \cref{eq:prob1} implies $p+q \le \frac{1}{\sqrt{2}}$. This means $r + s + t \ge 1 - \frac{1}{\sqrt{2}}$. Since we assume $\beta \ge 2 + \sqrt{2}$, it is easy to check that the objective is strictly negative if $B \ge 100$. Therefore, add $0 \le B \le 100$ as a constraint. We find the global optimum via BARON~\cite{Sahinidis1996,KS18} for both cases. For $\theta_2 = \sqrt{2} - 1$, and $\beta = 2 + \sqrt{2} + \delta$ for small $\delta > 0$, the global optimum for both cases is strictly negative. This shows that $\beta = 2 + \sqrt{2}$, and the distortion is at most $1+\beta = 3 + \sqrt{2}$, which shows the the upper bound in \cref{thm:theta2}.

\paragraph{Matching Lower Bound.} The above program also unearths the worst case example that shows the bound is tight for rules that choose from the uncovered set.  Place $W,X,Y$ on a line in that order with a distance of $1$ between $W,X$ and $1$ between $X,Y$. Place a mass $1-\alpha$ of voters very close to  $X$ and $\alpha$ very close to $Y$, where $\alpha = 1 - 1/\sqrt{2}$. The distances can be appropriately set so that the following happens under favorable tie-breaking: Whenever both voters in a sampled pair are located close to $X$ (which happens with probability $(1-\alpha)^2 = 1/2$), they will prefer $W$ over $Y$, so that $p_k(W,Y) \ge 1/2$. Next, when one voter is close to $X$ and the other is close to $Y$, they prefer $Y$ to $X$. This means $p_k(Y,X) = 1 - (1-\alpha)^2 = 1/2$. This means $W$ is in the uncovered set. The social optimum is $X$, and the distortion is $\frac{SC(W)}{SC(X)} = 1+1/\alpha = 3 + \sqrt{2}$. This completes the proof of \cref{thm:theta2}.

%% file: average_three.tex
\subsection{Distortion Bound for $k=3$}
We show that groups of size $3$ suffice for the distortion to go below the deterministic metric distortion lower bound of $3$. (As noted before, the Averaging rule for a finite but large number of voters can be viewed as a deterministic social choice rule.) Note that for $k=3$, the proof of \cref{thm:lb1} implies $\theta_3 \ge 0.25$, which also implies a lower bound of $5/3 \approx 1.667$ on the distortion of any deterministic social choice rule that uses outcomes of all deliberation triplets (\cref{thm:lb_main}). We show that the bound on $\theta_3$ is nearly tight.

\begin{theorem}
\label{thm:theta3}
For the optimization problem \cref{eq:opt_avg} with group size $k=3$, we have $\theta_3 \in [0.25, 0.2501]$.\footnote{The slack we introduce is due to numerical precision error of the global optimizer.} By \cref{thm:distort1}, this implies the distortion of the Copeland rule with $k = 3$ is at most $2.78$.
\end{theorem}

The rest of this subsection is devoted to proving $\theta_3 \le 0.2522$. We write a compact non-linear programming relaxation for the optimum of \cref{eq:opt_avg}. Note that the optimal distribution has expectation $\theta_3$. 

The key to our relaxation is the following lemma, which is a corollary of Lemma 1 in~\cite{Feige}. We present a proof for completeness.

\begin{lemma}[\cite{Feige}]  There exist independent distributions $D_1, D_2, D_3$ each supported on two values in $[-1,1]$ (these values could be different for each $D_i$), such that $\E[D_1] = \E[D_2] = \E[D_3] = \theta_3$, and 
\begin{equation}
\label{eq:feige}
\Pr_{a_1  \sim D_1, a_2 \sim D_2, a_3 \sim D_3} \left[ a_1 + a_2 + a_3 \le 0 \right] \ge 1/2. 
\end{equation}
\label{lem:feige}
\end{lemma}
\begin{proof}
We start with the distribution $D^*$ that optimizes \cref{eq:opt_avg}. Let $Y_1,Y_2,Y_3$ denote $i.i.d.$ random variables with distribution $D^*$ so that $a_1 \sim Y_1$, $a_2 \sim Y_2$, and $a_3 \sim Y_3$. First consider $Y_1$, and assume w.l.o.g. that it is finely discretized as $\{v_1,v_2, \ldots, v_n\}$, and let $p_j = \Pr[Y_1 = v_j]$. Let $q_j = \Pr[Y_2 + Y_3 + v_j \le 0]$. Then
$$ \sum_{j=1}^n p_j = 1 \qquad \mbox{and} \qquad \E[D^*] = \sum_{j=1}^n p_j v_j \qquad \mbox{and} \qquad \Pr[Y_1 + Y_2 + Y_3 \le 0] = \sum_{j=1}^n p_j q_j \ge 1/2.$$
Now treat the $p_j$ as non-negative variables, and maximize the LHS of the third constraint subject to the first two constraints. The optimum has at most two non-zero variables, and satisfies the third constraint (since the LHS cannot decrease). This yields a new distribution $D_1$ with $\E[D_1] = \theta_3$, and with support two. Repeat this process for $Y_2$ and then $Y_3$, completing the proof.
\end{proof}

We will find the optimal $(D_1,D_2, D_3)$ satisfying the above lemma, and this will upper bound $\theta_3$. For $i = 1,2,3$, let $D_i = a_i$ with probability $1-p_i$ and $b_i$ with probability $p_i$, where for all $i = 1,2,3$. Therefore, we have the constraints:
\begin{equation}
\label{eq:3.1}
    b_i \le a_i; \qquad a_i, b_i \in [-1,1],  \qquad  c_i = a_i - b_i, \qquad p_i \in [0,1] \qquad \forall i = 1,2,3. 
\end{equation} 
The $\{a_i,b_i,c_i,p_i\}_{i=1}^3$ and $\theta_3$ will be the variables in our program. We also have the constraints:
\begin{equation} 
\label{eq:3.2}
\theta_3 = b_i p_i + a_i (1-p_i) \qquad \forall i = 1,2,3. 
\end{equation}
Our goal is to maximize $\theta_3$. W.l.o.g., enforce the constraint that 
\begin{equation}
\label{eq:3.3}
c_3 \ge c_2 \ge c_1 \ge 0.
\end{equation}
Write $D_i = a_i - Z_i$ where $Z_i = c_i$ with probability $p_i$ and $0$ otherwise. Then the constraint in \cref{eq:feige} translates to:
\begin{equation}
\label{eq:zee}
\Pr_{z_1  \sim Z_1, z_2 \sim Z_2, z_3 \sim Z_3} \left[ \sum_{i=1}^3 z_i\ge \sum_{i=1}^3 a_i \right] \ge 1/2. 
\end{equation}

\begin{table*}[htbp]
\centering
\rowcolors{2}{gray!25}{white}
\begin{tabular}{|c|l|l|}
\hline 
Case &  Constraint on range of $\sum_{i=1}^3 a_i$ & Constraint \cref{eq:zee} \\
\hline 
1 & $ c_3 + c_2 + c_1 \ge a_3 + a_2 + a_1 \ge c_3 + c_2$ & $ p_1 p_2 p_3 \ge 1/2$ \\
2 &  $ c_3 + c_2  \ge a_3 + a_2 + a_1 \ge c_3 + c_1$ & $p_3 p_2 \ge 1/2$ \\
3 & $ c_3 + c_1  \ge a_3 + a_2 + a_1 \ge \max\{c_3, c_2 + c_1\}$  & $p_3 p_2 + p_3 p_1 (1-p_2) \ge 1/2.$ \\
4 &  $c_3   \ge a_3 + a_2 + a_1 \ge c_2 + c_1$ & $ p_3  \ge 1/2.$ \\
5 &  $ c_2+c_1   \ge a_3 + a_2 + a_1 \ge c_3 $ & $p_3  p_2   + p_3  (1-p_2)  p_1 + (1-p_3)  p_2 p_1 \ge 1/2$ \\
6 & $ \min\{c_2+c_1,c_3\}   \ge a_3 + a_2 + a_1 \ge c_2 $ & $ 1 - (1-p_3)  (1 - p_1 p_2) \ge 1/2$ \\
7 & $ c_2   \ge a_3 + a_2 + a_1 \ge c_1 $ & $ 1 - (1-p_3)  (1-p_2) \ge 1/2$ \\
8 &  $c_1 \ge a_3 + a_2 + a_1 $ & $ 1 - (1-p_3)  (1-p_2) (1-p_1) \ge 1/2$ \\
\hline
\end{tabular}
\caption{\label{tab:cases} The possible ranges for $\sum_{i=1}^3 a_i$ and the encoded constraints for each case.}
\end{table*}

In sorted order, $\sum_{i=1}^3 z_i$ can take values 
$$ c_3 + c_2 + c_1 \ge c_3 + c_2 \ge c_3 + c_1 \ge \{c_2 + c_1, c_3 \} \ge c_2 \ge c_1 \ge 0.$$
We enforce the constraint that the quantity $\sum_{i=1}^3 a_i$ lies between two of the values; this yields eight possible programs. In each of these programs, we have the constraints \cref{eq:3.1,eq:3.2,eq:3.3} and the objective is to maximize $\theta_3$. The cases differ in how \cref{eq:zee} is encoded. These cases are shown in \cref{tab:cases}, where the constraint in the third column encodes \cref{eq:zee} when $\sum_{i=1}^3 a_i$ lies in the range encoded by the constraint in the second column. One end-point in the second column should be a strict inequality, but relaxing it to a weak inequality only makes the corresponding objective used for our upper bound weakly larger. 

As an example, for the first row, we have $\sum_{i=1}^3 a_i \in [c_3 + c_2, c_3 + c_2 + c_1]$. But this means for the event in \cref{eq:zee} to be satisfied, we must have $z_3 = c_3, z_2 = c_2, z_1 = c_1$, which happens with probability $p_3 p_2 p_1$. The constraint \cref{eq:zee} therefore implies $p_1 p_2 p_3 \ge 1/2$. We encode the constraints in the second and third columns of the row corresponding to Case (1) into the program for Case (1), and solve it along with constraints \cref{eq:3.1,eq:3.2,eq:3.3}, and the objective of maximizing $\theta_3$.

For each of the eight cases, we globally maximize $\theta_3$ using BARON~\cite{Sahinidis1996,KS18}. We observe that  the global optimum is found to be at most $0.2501$.  Plugging this into \cref{thm:distort1} completes the proof of \cref{thm:theta3}. 


%% file: average_sample.tex
\subsection{Sample Complexity}
So far, our bounds have assumed $p_k(W,X)$ --- the probability that a randomly sampled group of size $k$ leads to $W$ as the outcome of deliberation --- can be estimated exactly. Via a standard argument, this can be converted to a sampling bound on the number of sampled groups needed to estimate $p_k(W,X)$ for all pairs of alternatives.



For $\eta \ge 0$, define the relaxed version of \cref{eq:opt_avg} as
$$
\theta_k(\eta) := \sup_D \left\{ \E[D] : D \mbox{ supported on } [-1,1],\ 
\Pr_{a_1,\ldots,a_k \sim D}\left[\sum_{i=1}^k a_i \le 0\right] \ge \frac{1}{2}-\eta \right\}.
$$
Thus $\theta_k(0)=\theta_k$. We first note that, for every fixed group size $k$, this relaxation changes the optimum continuously at a linear rate.

\begin{lemma}
\label{lem:theta_smooth}
Fix $k \ge 2$. For any $\eta \in [0,1/2]$, we have
$\theta_k(\eta) \le \theta_k + O_k(\eta)$.
\end{lemma}

\begin{proof}
Let $D$ be any distribution feasible for $\theta_k(\eta)$, and write
$H(D)=\Pr_{a_1,\ldots,a_k\sim D}[\sum_i a_i \le 0]$ and $\mu=\E[D]$.
Here $D$ denotes a probability distribution on $[-1,1]$.
If $H(D)\ge 1/2$, then $D$ is feasible for $\theta_k$, so $\mu\le \theta_k$a nd there is also nothing to prove. We therefore assume
$1/2-\eta \le H(D)<1/2$ and $\mu>\theta_k$. Since $\theta_k>0$ by \cref{thm:lb1}, 
we have $\mu>0$.

We will slightly reweight $D$ so that values below the mean receive more
probability and values above the mean receive less probability. For
$t\in[0,1/2]$, define a new distribution $D_t$ by declaring that, for every
measurable set $A\subseteq[-1,1]$,
$$
D_t(A)=\int_A \bigl(1+t(\mu-x)\bigr)\,dD(x).
$$
Equivalently, $D_t$ is the distribution obtained from $D$ by multiplying the
probability mass near $x$ by the factor $1+t(\mu-x)$. Thus, if $D$ has a density $f$, then $D_t$ has density $f_t(x)=f(x)(1+t(\mu-x))$. 

We first check that $D_t$ is a probability distribution. Since
$x,\mu\in[-1,1]$, we have $\mu-x\ge -2$, and therefore
$1+t(\mu-x)\ge 1-2t\ge 0$. Moreover,
$$
D_t([-1,1])
=\int_{-1}^1 \bigl(1+t(\mu-x)\bigr)\,dD(x)
=1+t(\mu-\E[D])
=1.
$$
Therefore $D_t$ is a probability distribution supported on $[-1,1]$.

The expectation of $D_t$ is
$$
\E[D_t]
=\int_{-1}^1 x\bigl(1+t(\mu-x)\bigr)\,dD(x)
=\mu+t(\mu\E[D]-\E[D^2])
=\mu-t(\E[D^2]-\mu^2).
$$
In other words, $\E[D_t]=\mu-t\mathrm{Var}(D)$, where
$\mathrm{Var}(D)=\E[D^2]-\mu^2$. Thus this reweighting decreases the
expectation by $t\mathrm{Var}(D)$.

We now show that $H(D_t)$ increases linearly in $t$. Draw
$a_1,\ldots,a_k$ independently from $D$, and write $S=\sum_i a_i$. Since
$D_t$ is obtained from $D$ by the reweighting factor $1+t(\mu-x)$, the product
distribution $D_t^k$ is obtained from $D^k$ by multiplying the probability of
the tuple $(a_1,\ldots,a_k)$ by
$\prod_{i=1}^k (1+t(\mu-a_i))$. Hence
$$
H(D_t)
=\E_{a_1,\ldots,a_k\sim D}
\left[
{\bf 1}_{\{S\le 0\}}
\prod_{i=1}^k \bigl(1+t(\mu-a_i)\bigr)
\right].
$$
Let $u_i=\mu-a_i$. Since $\mu,a_i\in[-1,1]$, we have $|u_i|\le 2$. Expanding
the product, write
$\prod_{i=1}^k(1+tu_i)=1+t\sum_i u_i+R_t$, where $R_t$ contains all terms
of degree at least two in $t$. For $t\le 1/2$,
$$
|R_t|
\le \sum_{j=2}^k \binom{k}{j}(2t)^j
\le 4t^2\sum_{j=2}^k \binom{k}{j}
=4(2^k-k-1)t^2.
$$
Let $B_k=4(2^k-k-1)$. Substituting the expansion into the expression for
$H(D_t)$ and using $H(D)  = \E\left[
{\bf 1}_{\{S\le 0\}}\right]$, we have
$$
H(D_t)
\ge H(D)
+t\,\E\left[
{\bf 1}_{\{S\le 0\}}
\sum_i(\mu-a_i)
\right]
-B_kt^2.
$$
On the event $S\le 0$, we have $\sum_i(\mu-a_i)=k\mu-S\ge k\mu$. Therefore, we have
$$
H(D_t)\ge H(D)+k\mu t \cdot H(D)-B_kt^2.
$$

We first prove the claim for sufficiently small $\eta$. Suppose $\eta\le 1/4$.
Since $H(D)\ge 1/2-\eta$ and $\mu>\theta_k$, we have $H(D)\ge 1/4$, and hence
$k\mu H(D)\ge k\theta_k/4$. Let $c_k=k\theta_k/4  >0$. Note $c_k = \Theta(1)$ by \cref{sec:berry}. Then we have:
$$
H(D_t)\ge H(D)+c_kt-B_kt^2.
$$
Set $t=2\eta/c_k=8\eta/(k\theta_k)$. Since $c_k>0$ and $B_k>0$ are constants depending only on $k$, choose
$
\eta_0(k)=\min\left\{\frac14,\frac{c_k}{4},\frac{c_k^2}{4B_k}\right\}.
$
We claim that whenever $\eta\le \eta_0(k)$, we have $\eta\le 1/4$, $t\le 1/2$, and $B_kt^2\le \eta$. Indeed, the first part is direct. Also, since
$t=2\eta/c_k$ and $\eta\le c_k/4$, we have $t \le \frac12.$
Finally, since $\eta\le c_k^2/(4B_k)$, we have
$$
B_kt^2
=
B_k\left(\frac{2\eta}{c_k}\right)^2
=
\frac{4B_k\eta^2}{c_k^2}
\le \eta.
$$ For this choice of $t$,
$$
H(D_t)
\ge H(D)+c_kt-B_kt^2
\ge H(D)+2\eta-\eta
\ge \frac12.
$$
Thus $D_t$ is feasible for $\theta_k$, so $\E[D_t]\le \theta_k$. Using
$\E[D_t]=\mu-t\mathrm{Var}(D)$ and $\mathrm{Var}(D)\le 1$, we get
$$
\mu
=\E[D_t]+t\mathrm{Var}(D)
\le \theta_k+t
=\theta_k+\frac{8}{k\theta_k}\eta.
$$
Taking the supremum over all distributions $D$ feasible for $\theta_k(\eta)$
proves $\theta_k(\eta)\le \theta_k+O_k(\eta)$ whenever $\eta\le \eta_0(k)$.

It remains to extend the bound from $\eta\le\eta_0(k)$ to all
$\eta\in[0,1/2]$. From the previous paragraph, there is a constant
$C_0(k)$ such that
$\theta_k(\eta)\le \theta_k+C_0(k)\eta$ whenever $\eta\le\eta_0(k)$.
Now suppose $\eta>\eta_0(k)$. Note that
$\theta_k(\eta)\le 1$ and
$\eta/\eta_0(k)>1$. Since $\theta_k\ge0$, we get
$$
\theta_k(\eta)\le 1\le \theta_k+\frac{\eta}{\eta_0(k)}.
$$
Thus the desired inequality also holds for $\eta>\eta_0(k)$, with constant
$1/\eta_0(k)$, completing the proof for all
$\eta\in[0,1/2]$.
\end{proof}

For each sampled group of size $k$, we ask the voters to rank all alternatives.
Assuming that $c_1$ is ranked higher than $c_2$ if
$\sum_{i\in S} d(i,c_1)\le \sum_{i\in S} d(i,c_2)$, it is easy to check that for
each pair $(W,X)$, the outcome of deliberation just among this pair of alternatives
is consistent with the ranking. For a pair $(W,X)$, let $\widehat p_k(W,X)$ denote
the fraction of sampled groups that rank $W$ above $X$. By a standard application
of Chernoff bounds~\cite{Boucheron}, if we sample
$O(\eta^{-2}\log(m/\delta))$ groups of size $k$, then with probability at least
$1-\delta$, every $p_k(W,X)$ is approximated by $\widehat p_k(W,X)$ to within
additive error $\eta$.

Let $W$ be the Copeland winner in the tournament defined by the empirical
probabilities $\widehat p_k$. On the high-probability event above, for every
alternative $X$, either $p_k(W,X)\ge 1/2-\eta$, or there exists an alternative
$Y$ such that $p_k(W,Y)\ge 1/2-\eta$ and $p_k(Y,X)\ge 1/2-\eta$. Indeed, the
corresponding statement holds in the empirical tournament because a Copeland
winner is in the uncovered set, and each empirical comparison with value at
least $1/2$ translates to a true comparison with value at least $1/2-\eta$.

Now consider any comparison $A$ versus $B$ with $p_k(A,B)\ge 1/2-\eta$. In the
averaging model, if $D$ is the distribution of normalized biases $B_i(A,B)$,
then this condition implies the feasibility condition in the definition of
$\theta_k(\eta)$. Therefore $\E[D]\le \theta_k(\eta)$. By the proof of \cref{thm:distort1},
the Copeland winner computed from the empirical tournament has distortion at most
$
\left(\frac{1+\theta_k(\eta)}{1-\theta_k(\eta)}\right)^2.$
By \cref{lem:theta_smooth}, $\theta_k(\eta)\le \theta_k+O_k(\eta)$.
Choose $\eta_0=\eta_0(k)>0$ small enough that $\theta_k(\eta)\le
(1+\theta_k)/2$ for all $\eta\le \eta_0$. Since the function
$x\mapsto ((1+x)/(1-x))^2$ has bounded derivative on
$[\theta_k,(1+\theta_k)/2]$, for all $\eta\le \eta_0$ the above distortion is at most
$
\left(\frac{1+\theta_k}{1-\theta_k}\right)^2
+O_k\left(\eta\right).
$
This yields the following sampling theorem.

\begin{theorem}
\label{thm:sample2}
Fix $k\ge 2$ and let
$d_k=\left(\frac{1+\theta_k}{1-\theta_k}\right)^2$ denote the distortion bound
obtained by applying \cref{thm:distort1} to \cref{eq:opt_avg}. If $0<\eta\le \eta_0(k)$ and
$N=O(\eta^{-2}\log(m/\delta))$ groups of size $k$ are sampled, then the Copeland
winner of the empirical tournament has distortion at most
$d_k+O_k(\eta)$ with probability $1-\delta$. 
\end{theorem}

As shown in \cref{eg1},  if the group size $k$ is a constant and we want to beat the distortion bound of $3$, then this crucially requires the deliberating group to rank alternatives {\em beyond} their favorite (or median) alternative (among all $m$ alternatives).  This makes our sampling bound above and in \cref{thm:sample1} with constant size groups different from core-set type sampling bounds for the $1$-median problem, for instance, the bounds in~\cite{CaragiannisM024}. 

%% file: random.tex
\section{Distortion in the Random Choice Model}
\label{sec:random}
We next compute the distortion in the Random Choice model for various values of group size $k$. In contrast to the averaging model, this model is more analytically tractable, and we can numerically compute a good upper bound on distortion for all small $k$. In particular, our main result in \cref{thm:random} is that groups of size $k = 4$ suffice to break the randomized metric distortion lower bound of $2.11$. In \cref{thm:asymp1}, we also show that the group size is $k = \tilde{O}(1/\epsilon^2)$ for Copeland to achieve distortion $1+\epsilon$. We also show in \cref{thm:sample1} that a sample of $\tilde{O}(m \log m)$ groups suffices for approximating distortion, where $m$ is the number of alternatives.  In \cref{sec:general}, we finally present a generalization of the random choice model, and show how our analysis technique naturally extends to showing distortion bounds for it.

\subsection{Distortion Bounds for Small $k$}
\label{sec:random_small}
Suppose the deliberation has $k$ participants, and is between two arbitrary outcomes $W$ and $X$. For a distribution $D$, let $L$ be the conditional distribution of $-D$ given $D \le 0$, and let $R$ denote the conditional distribution of $D$ given $D > 0$. Let $\alpha = \Pr[D \le 0]$. The objective in \cref{eq:opt} can be written as
$$ \max \ \ (1-\alpha) \cdot \E[R] - \alpha \cdot \E[L].$$

Let $a_1, a_2 \ldots, a_k$ be $k$ i.i.d. samples from $L$ and $b_1, b_2 \ldots, b_k$ be $k$ i.i.d. samples from $R$. Then,
$$ p_k(W,X) = \sum_{\ell = 1}^k {k \choose \ell} \alpha^{\ell} (1-\alpha)^{k-\ell} \E\left[ \frac{\sum_{r = 1}^{\ell} a_r}{\sum_{r=1}^{\ell} a_r + \sum_{q=1}^{k-\ell} b_q } \right],$$
where the expectation is over $a_1,\ldots, a_r \overset{\text{i.i.d.}}{\sim} L$, and $b_1,\ldots, b_q \overset{\text{i.i.d.}}{\sim}  R$. The constraint in \cref{eq:opt} implies the RHS is at least $1/2$. 
Note now that since the RHS is convex in any $b_q$, we can preserve the objective above and increase the RHS when $b_q$ is drawn from a Bernoulli distribution with mean $\E[R]$. We next absorb the mass at $0$ in $R$ into the distribution $L$; call the new distributions $\hat{L}, \hat{R}$. Therefore, $\hat{R}$ is the deterministic value $1$, and $\Pr[\hat{R}] = 1-\alpha$. Then, the objective is
$$ (1-\alpha) - \alpha \E[\hat{L}].$$
Since each $b_q = 1$ now, the constraint on $p_k(W,X) $  becomes
$$ p_k(W,X) \le \sum_{\ell = 1}^k {k \choose \ell} \alpha^{\ell} (1-\alpha)^{k-\ell} \E\left[ \frac{\sum_{r = 1}^{\ell} a_r}{\sum_{r=1}^{\ell} a_r + k-\ell } \right],$$
where the expectation is now over $a_1,\ldots,a_k\overset{\text{i.i.d.}}{\sim} \hat{L}$. Noting that the RHS is concave in $a_r$, by Jensen's inequality, we have
$$ p_k(W,X) \le \sum_{\ell = 1}^k {k \choose \ell} \alpha^{\ell} (1-\alpha)^{k-\ell} \cdot \frac{\ell \cdot \E[\hat{L}]}{\ell \cdot \E[\hat{L}] + k-\ell }. $$

Therefore, using $\omega = \E[\hat{L}]$ and observing that $\omega, \alpha \in [0,1]$, \cref{eq:opt} can be rewritten as:
\begin{equation} 
\label{eq:opt2}
\zeta_k := \max_{\omega, \alpha \in [0,1]} (1-\alpha) - \alpha \cdot \omega \qquad \mbox{s.t.} \qquad 
\sum_{\ell = 1}^k {k \choose \ell} \alpha^{\ell} (1-\alpha)^{k-\ell} \cdot
\left[ \frac{\ell \cdot \omega}{\ell \cdot \omega + k-\ell } \right] \ge \frac{1}{2}.
\end{equation}

\begin{figure*}[htbp]
    \centering
    \begin{subfigure}[t]{0.4\textwidth}
        \centering
        \includegraphics[width=\linewidth]{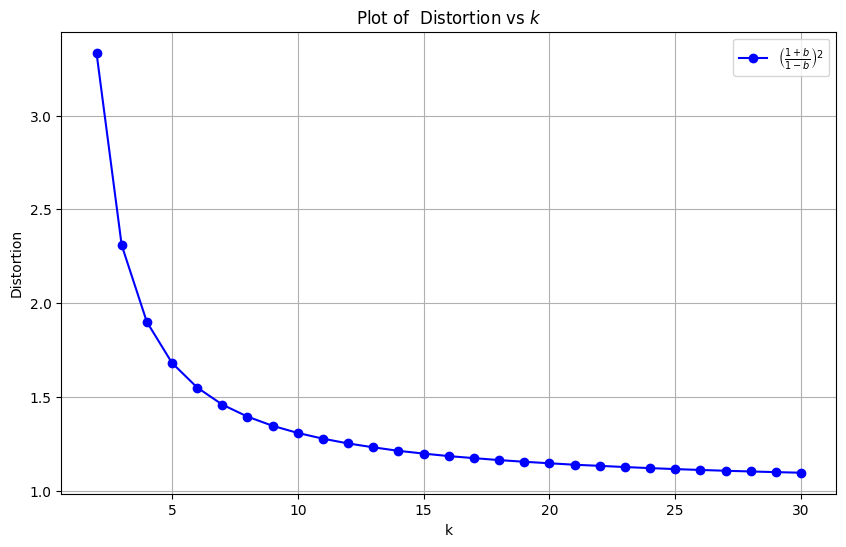}
        \caption{\label{fig1} Random Choice Model.}
    \end{subfigure}
    \qquad
        \begin{subfigure}[t]{0.4\textwidth}
        \centering
        \includegraphics[width=\linewidth]{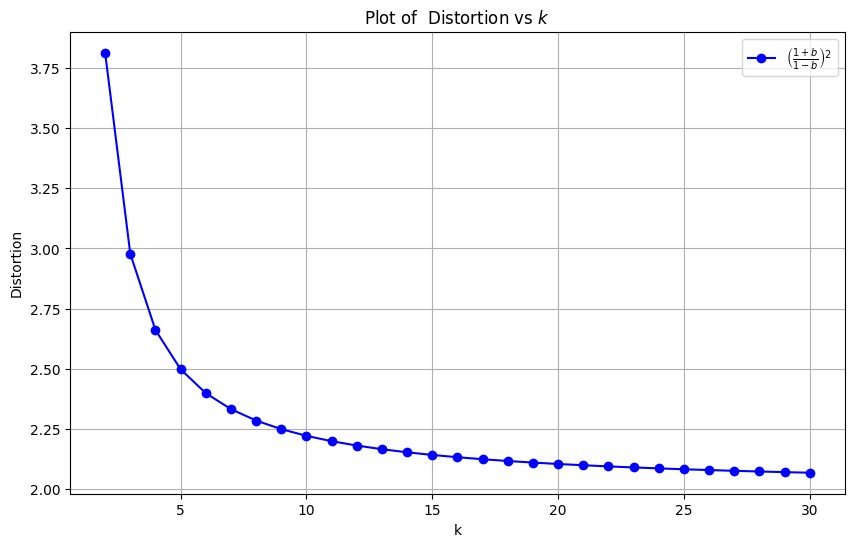}
        \caption{\label{fig2} Generalized Random Choice, $g(x) = \sqrt{x}$.}
    \end{subfigure}
    \caption{Distortion of the Random Choice model as a function of the group size $k$, for $k \in [2,30]$.}
\end{figure*}

For any given $\alpha$, the LHS of the constraint is concave and increasing in $\omega$, while the objective is decreasing in it, so it can be optimized by a binary search over $\omega$ to find the smallest $\omega$ for which the constraint is satisfied. We then run a parametric search over $\alpha$ in increments of $10^{-3}$ to find an approximate global optimum. By \cref{thm:distort1}, the distortion is at most $\left(\frac{1+\zeta_k}{1-\zeta_k}\right)^2$. We numerically compute this for $2 \le k \le 30$, and plot it in \cref{fig1}, and show in the second column in \cref{tab:random}. 

\paragraph{Rational Certificate of Bounds.} We now formally certify an upper envelope for the objective for $k = 2,3,4$ without relying on the grid search.  For fixed $k$, let $
F_k(\alpha,\omega)=
\mathbb{E}_{\ell\sim\mathrm{Binomial}(k,\alpha)}
\left[
\frac{\ell\omega}{\ell\omega+k-\ell}
\right],
$
 Let
$
g(\alpha,\omega)=(1-\alpha)-\alpha\omega.
$
The constraint in \cref{eq:opt2} is $F_k(\alpha,\omega)\ge 1/2$, and the objective is
$g(\alpha,\omega)$.

First note that no feasible solution has $\alpha<1/2$, since
$F_k(\alpha,1)=\alpha$.  We therefore restrict attention to
$\alpha\in[1/2,1]$.  The function $F_k$ is nondecreasing in $\omega$.  It is also
nondecreasing in $\alpha$, since the summand is nondecreasing in $\ell$ and the
binomial distribution is stochastically increasing in $\alpha$.

Partition $[1/2,1]$ into finitely many intervals $I_j=[a_j,b_j]$.  For each interval, our certification code computes a rational number $\underline{\omega}_j$ such that
$
F_k(b_j,\underline{\omega}_j)\le \frac12.
$
If
$F_k(b_j,0) >  1/2$, we set $\underline{\omega}_j=0$. Therefore, if $(\alpha,\omega)$ is feasible and $\alpha\in I_j$, monotonicity
implies $\omega\ge \underline{\omega}_j$.  Hence
\[
g(\alpha,\omega)
=1-\alpha(1+\omega)
\le 1-a_j(1+\underline{\omega}_j).
\]
It follows that
\[
\zeta_k
\le
\max_j \left\{1-a_j(1+\underline{\omega}_j)\right\}.
\]
All quantities in this maximization are rational, and the code
performs the comparisons exactly.  The certified upper bounds are
\[
\zeta_2 \le 0.293274819,\qquad
\zeta_3 \le 0.206306458,\qquad
\zeta_4 \le 0.16015625.
\]
Plugging these into $\left(\frac{1+\zeta_k}{1-\zeta_k}\right)^2$ gives
distortion at most $3.35$ for $k=2$, at most $2.31$ for $k=3$, and at most
$1.91$ for $k=4$. This yields the following theorem.

\begin{theorem}
\label{thm:random}
    For the Copeland rule applied to the random choice deliberation model with group size $k$, the distortion is at most $3.35$ for $k = 2$, at most $2.31$ for $k = 3$, and at most $1.91$ for $k = 4$.
\end{theorem}

Note that the upper bound for $k = 2$ is significantly lower than the lower bound of $4.414$ for the averaging model in \cref{thm:theta2}. Note further that the distortion for $k = 3$ is well below the distortion of $2.74$ of the best randomized social choice rule without deliberation, and that for $k = 4$ is below the lower bound of $2.11$ on metric distortion without deliberation. 

\paragraph{Lower Bound for any  Social Choice Rule.} 
Let $\zeta_k$ be the optimal value to \cref{eq:opt2} for groups of size $k$. Note that this corresponds to the expectation of a valid distribution over $\{-\omega, 1\}$, where $\Pr[-\omega] = \alpha$. By \cref{thm:lb_main}, this implies a distortion of $\frac{1+\zeta_k}{1-\zeta_k}$ for any deterministic social choice rule, and $\frac{1}{1-\zeta_k}$ for any randomized rule. These values are shown in the third and fourth columns in \cref{tab:random}, and show that our upper bounds above are reasonably tight.  

%% file: random_asymptotic.tex
\subsection{Asymptotic Behavior of Distortion in Group Size $k$}
We next show that the group size needed for Copeland to achieve a distortion of $1+\epsilon$ is $\tilde{O}(1/\epsilon^2)$, independent of the number of alternatives.

\begin{theorem}
\label{thm:asymp1}
For any $\epsilon > 0$, with a group size of $k = O\left(\frac{1}{\epsilon^2} \log\frac{1}{\epsilon}\right)$, the distortion of the Copeland rule in the random choice deliberation model is at most $1+\epsilon$.
\end{theorem}

\begin{proof}
Consider the optimization problem in \cref{eq:opt2}. Let
$
F_k(\alpha,\omega)
=
\mathbb{E}_{\ell\sim \mathrm{Binomial}(k,\alpha)}
\left[
\frac{\ell\omega}{\ell\omega+k-\ell}
\right],
$
where the summand is interpreted as $1$ when $\ell=k$. The objective is
$
g(\alpha,\omega)=(1-\alpha)-\alpha\omega.
$

If $\alpha<1/2$, then
$
F_k(\alpha,1)=\mathbb{E}[\ell/k]=\alpha<1/2,
$. Since $F_k(\alpha,\omega)$ is monotone in $\omega$, this means 
 no feasible solution has $\alpha<1/2$. Fix $0<\delta\le 1/2$, and suppose
$
k\ge \frac{6}{\delta^2}\log\frac{4}{\delta}.
$
Let $(\alpha,\omega)$ be feasible, so $\alpha\ge 1/2$. If
$\alpha\ge 1/(1+\delta)$, then
\[
g(\alpha,\omega)\le 1-\alpha\le \frac{\delta}{1+\delta}\le \delta.
\]
Assume $\alpha<1/(1+\delta)$. Let $\ell\sim \mathrm{Binomial}(k,\alpha)$ and set
$
\beta=(1+\delta)\alpha<1.
$ By Chernoff bounds,
\[
\Pr[\ell\ge (1+\delta)k\alpha]
\le \exp\left(-\frac{k\alpha\delta^2}{3}\right)
\le \exp\left(-\frac{k\delta^2}{6}\right)
\le \frac{\delta}{4}.
\]
The function
$
x\mapsto \frac{x\omega}{x\omega+k-x}
$
is increasing in $x$. Therefore, conditioning on whether
$\ell\ge (1+\delta)k\alpha$, feasibility implies
\[
\frac{\beta\omega}{\beta\omega+1-\beta}+\frac{\delta}{4}\ge \frac12 \qquad \Rightarrow \qquad
\frac{\beta\omega}{\beta\omega+1-\beta}\ge \frac12-\frac{\delta}{4}.
\]
Rearranging, and using $\delta\le 1/2$, gives
\[
\beta\omega
\ge
\frac{1-\delta/2}{1+\delta/2}(1-\beta)
\ge
(1-\delta)(1-\beta).
\]
Since $\beta=(1+\delta)\alpha$, we obtain
\[
\alpha\omega
=
\frac{\beta\omega}{1+\delta}
\ge
\frac{1-\delta}{1+\delta}\bigl(1-(1+\delta)\alpha\bigr).
\]
Therefore
\[
\begin{aligned}
g(\alpha,\omega)
&=1-\alpha-\alpha\omega
\le
1-\alpha-
\frac{1-\delta}{1+\delta}\bigl(1-(1+\delta)\alpha\bigr)=
\frac{2\delta}{1+\delta}-\delta\alpha
\le 2\delta.
\end{aligned}
\]
Thus $\zeta_k\le 2\delta$ whenever
$k=O(\delta^{-2}\log(1/\delta))$. Since
$
\left(\frac{1+\zeta_k}{1-\zeta_k}\right)^2=1+O(\zeta_k)
$
for $\zeta_k$ bounded away from $1$, choosing $\delta=\Theta(\varepsilon)$ and
applying \cref{thm:distort1} gives distortion at most $1+\varepsilon$ for
$
k=O\left(\frac{1}{\varepsilon^2}\log\frac{1}{\varepsilon}\right).
$
\end{proof}


%% file: random_sample.tex
\subsection{Sample Complexity}

As in \cref{thm:sample2}, we bound the number of samples needed to estimate all
$p_k(W,X)$ to within a small additive error. The key difference is that the
Random Choice model is only defined for pairs of alternatives. We first record
a stability property of the program in \cref{eq:opt2}. For $\eta\ge 0$, let
$\zeta_k(\eta)$ denote the optimum of the program in \cref{eq:opt2} after
replacing the right hand side of the constraint by $1/2-\eta$. Thus
$\zeta_k(0)=\zeta_k$.

\begin{lemma}
\label{lem:zeta_smooth}
For every fixed $k\ge 2$ and every $0\le \eta\le 1/4$, we have
$\zeta_k(\eta)\le \zeta_k+O_k(\eta)$.
\end{lemma}

\begin{proof}
For $\alpha,\omega\in[0,1]$, write
$$
F_k(\alpha,\omega)=
\mathbb{E}_{\ell\sim \mathrm{Binomial}(k,\alpha)}
\left[h_\ell(\omega)\right],
$$
where $h_\ell(\omega)=\frac{\ell\omega}{\ell\omega+k-\ell}$ for
$0\le \ell<k$, and $h_k(\omega)=1$. This is the left hand side of the
constraint in \cref{eq:opt2}. Let $g(\alpha,\omega)=(1-\alpha)-\alpha\omega$
denote the objective.

Consider any $(\alpha,\omega)$ feasible for $\zeta_k(\eta)$, so that
$F_k(\alpha,\omega)\ge 1/2-\eta$. If $F_k(\alpha,\omega)\ge 1/2$, then
$(\alpha,\omega)$ is feasible for $\zeta_k$, and hence
$g(\alpha,\omega)\le \zeta_k$. We therefore assume
$1/2-\eta\le F_k(\alpha,\omega)<1/2$.

Since $\eta\le 1/4$, we have $F_k(\alpha,\omega)\ge 1/4$. Also,
$$F_k(\alpha,\omega)\le \Pr[\mathrm{Binomial}(k,\alpha)\ge 1]
=1-(1-\alpha)^k,$$ since $h_0(\omega)=0$ and $h_\ell(\omega)\le 1$ for all
$\ell$. Combining these bounds, we have $\alpha\ge \underline{\alpha}_k$, where
$\underline{\alpha}_k=1-(3/4)^{1/k}> 0$ is a constant depending
only on $k$.

We now increase $\alpha$ while keeping $\omega$ fixed.  Using the standard identity, we have:
$$
\frac{\partial F_k}{\partial \alpha}
=
k\cdot
\mathbb{E}_{M\sim \mathrm{Binomial}(k-1,\alpha)}
\left[h_{M+1}(\omega)-h_M(\omega)\right].
$$
The sequence $h_0(\omega),h_1(\omega),\ldots,h_k(\omega)$ is nondecreasing. Keeping only the term $M=k-1$ in the
expectation, we get
$$\frac{\partial F_k}{\partial \alpha}\ge
k\alpha^{k-1}(h_k(\omega)-h_{k-1}(\omega)).$$ Since
$h_k(\omega)-h_{k-1}(\omega)=1/((k-1)\omega+1)\ge 1/k$, it follows that
$\frac{\partial F_k}{\partial \alpha}\ge \alpha^{k-1}\ge
\underline{\alpha}_k^{\,k-1}$ throughout the interval on which
$\alpha\ge \underline{\alpha}_k$.

Let $\alpha'\ge \alpha$ be the smallest value such that
$F_k(\alpha',\omega)\ge 1/2$; such a value exists because
$F_k(1,\omega)=1$. Since $F_k$ is continuous in $\alpha$, we have
$F_k(\alpha',\omega)=1/2$. Since $\alpha'\ge \alpha \ge \underline{\alpha}_k$, 
the derivative lower bound above applies throughout the interval $[\alpha,\alpha']$.
Using the choice of $\alpha'$ and the assumption $F_k(\alpha,\omega)\ge
1/2-\eta$, we get
$$
\eta \ge F_k(\alpha',\omega)-F_k(\alpha,\omega)
= \int_{\alpha}^{\alpha'}
\frac{\partial F_k}{\partial \alpha}(\beta,\omega)\,d\beta
\ge \underline{\alpha}_k^{\,k-1}(\alpha'-\alpha).
$$
Hence $\alpha'-\alpha\le \eta/\underline{\alpha}_k^{\,k-1}$. Since
$(\alpha',\omega)$ is feasible for $\zeta_k$, we have
$g(\alpha',\omega)\le \zeta_k$. Moreover,
$g(\alpha,\omega)=g(\alpha',\omega)+(1+\omega)(\alpha'-\alpha)$, so
$g(\alpha,\omega)\le
\zeta_k+2\eta/\underline{\alpha}_k^{\,k-1}=\zeta_k+O_k(\eta)$. Taking the
supremum over all feasible $(\alpha,\omega)$ proves the lemma.
\end{proof}

We now describe the sampling procedure. Consider the complete graph $K_m$ on
the alternatives, and decompose its edge set into $\chi\le m$ matchings. For
each matching, we sample $N_0$ independent groups of size $k$. A group assigned
to a matching deliberates over every pair of alternatives in that matching,
outputting one alternative for each pair using the Random Choice model. For a
pair $(W,X)$, let $\widehat p_k(W,X)$ denote the empirical fraction of sampled
groups assigned to its matching that output $W$ over $X$. By a standard Chernoff
bound and a union bound over all pairs of alternatives, if
$N_0=O(\eta^{-2}\log(m/\delta))$, then with probability at least $1-\delta$,
all quantities $p_k(W,X)$ are estimated by $\widehat p_k(W,X)$ to within
additive error $\eta$.

Condition on this high probability event, and let $W$ be the Copeland winner in
the empirical tournament defined by the probabilities $\widehat p_k$. Since a
Copeland winner lies in the uncovered set of the tournament, for every
alternative $X$, either $\widehat p_k(W,X)\ge 1/2$, or there exists an
alternative $Y$ such that $\widehat p_k(W,Y)\ge 1/2$ and
$\widehat p_k(Y,X)\ge 1/2$. Therefore, the same
statement holds for the true probabilities with threshold $1/2-\eta$: for every
$X$, either $p_k(W,X)\ge 1/2-\eta$, or there exists $Y$ such that
$p_k(W,Y)\ge 1/2-\eta$ and $p_k(Y,X)\ge 1/2-\eta$.

Now consider any comparison $A$ versus $B$ with $p_k(A,B)\ge 1/2-\eta$. The
same derivation as in \cref{eq:opt2}, with the right hand side of the constraint
replaced by $1/2-\eta$, implies that the corresponding value of
$\mathbb{E}[D]$ is at most $\zeta_k(\eta)$. Therefore, applying
the proof of \cref{thm:distort1} to the above relation gives
distortion at most
$
\left(\frac{1+\zeta_k(\eta)}{1-\zeta_k(\eta)}\right)^2.
$
By \cref{lem:zeta_smooth}, $\zeta_k(\eta)\le \zeta_k+O_k(\eta)$. Also, for
$\eta\le 1/4$, any feasible solution to $\zeta_k(\eta)$ has
$\alpha\ge 1-(3/4)^{1/k}$, and hence objective value at most $(3/4)^{1/k}<1$.
Thus the function $x\mapsto ((1+x)/(1-x))^2$ has derivative bounded by a
constant depending only on $k$ over the relevant range. Hence the above
distortion is at most
$$
\left(\frac{1+\zeta_k}{1-\zeta_k}\right)^2+O_k(\eta).
$$
This yields the following theorem.

\begin{theorem}
\label{thm:sample1}
Fix $k\ge 2$, and let
$d_k=\left(\frac{1+\zeta_k}{1-\zeta_k}\right)^2$ denote the distortion bound
obtained by applying \cref{thm:distort1} to \cref{eq:opt2}. Under the matching-based sampling procedure above, if
$N=O(m\eta^{-2}\log(m/\delta))$ groups of size $k$ are sampled, then the
Copeland winner of the empirical tournament has distortion at most
$d_k+O_k(\eta)$ with probability at least $1-\delta$. 
\end{theorem}

As mentioned after \cref{thm:sample2}, our sampling bounds with constant size groups are not implied by similar bounds for the $1$-median problem, and crucially require voters to output rankings beyond their favorite (or median) alternative.

%% file: random_general.tex
\subsection{Generalized Random Choice}
\label{sec:general}
We now consider two generalizations to the model that are amenable to the same analysis technique as in \cref{sec:random_small}. We present the model and the program in each case, and except for one canonical instance, we omit the easy computation of the numerical bounds for distortion. 

\paragraph{Concave Bias.} In the first generalization, the outcome of deliberation is less dominated by extremely biased voters. The model now has a concave, non-decreasing function $g$, with $g(0) = 0$ and $g(1) = 1$. For any pair of outcomes $W,X$, given a multiset $S$ of voter locations of size $k$, the probability the outcome of deliberation is the favorite outcome of $i \in S$ is proportional to $g(|\B_i(W,X)|)$. The Random Choice model considered so far had $g(x) = x$. Note that as $g$ becomes more concave (away from linear), the model favors a more uniformly random voter in the deliberating group, as opposed to a more biased voter.

For any such $g$, the function $\frac{g(x)}{g(x) + c}$ is also concave and non-decreasing in $x$. We can re-derive the optimization problem in \cref{eq:opt2} by analogously applying Jensen's inequality. This yields
\begin{equation} 
\label{eq:opt3}
\zeta_k := \max_{\omega, \alpha \in [0,1]} (1-\alpha) - \alpha \cdot \omega \qquad \mbox{s.t.} \qquad \E_{\ell \sim \mathtt{Binomial}(k,\alpha)} \left[ \frac{\ell \cdot g(\omega)}{\ell \cdot g(\omega) + k-\ell } \right] \ge \frac{1}{2}.
\end{equation}
for a distortion of at most $\left( \frac{1+\zeta_k}{1-\zeta_k}\right)^2$ by \cref{thm:distort1}. It is now easy to compute the distortion for any fixed function $g$. We plot the distortion as a function of $k$ for $g(x) = \sqrt{x}$ in \cref{fig2}. Observe that the distortion is at most $2.98$ for $k = 3$, and asymptotically approaches $2$ as $k \rightarrow \infty$.  

This shows that the asymptotic distortion bound of $1$ in \cref{thm:asymp1} is specific to the model with $g(x) = x$, and making $g(x)$ more concave leads to asymptotically larger distortion. 

\paragraph{Opinion Change of Voters.} In the second generalization, we model a voter in the group $S$ as changing their opinion. Let $S_1 \subseteq S$ denote the set of voters who prefer $W$ to $X$, and let $S_2 \subseteq S$ denote those that prefer $X$ to $W$. Let $A = \sum_{i \in S_1} |\B_i(W,X)|$ denote the total normalized bias of voters preferring $W$, and let $B = \sum_{i \in S_2} |\B_i(W,X)|$ be that for $X$. 

There is a parameter $\beta \in [0,1]$. During deliberation, suppose every voter in $S_1$ independently changes their opinion to $X$ with probability $\beta \cdot \frac{B}{A+B}$, and similarly, every voter in $S_2$ changes their opinion to $W$ with probability $\beta \cdot \frac{A}{A+B}$. Subsequently, the opinion of a randomly chosen voter in the group is implemented. It can be checked that
$$ p_k(W,X) = \beta \cdot \frac{A}{A+B} + (1-\beta) \cdot \frac{|S_1|}{|S|}.$$
Note that if $\beta = 1$, this is exactly the random choice model, whereas if $\beta = 0$, this is simply random dictatorship. The parameter $\beta$ captures the extent of opinion change. For any $\beta \in [0,1]$, we can upper bound the distortion using the same technique as in \cref{sec:random_small}. Indeed, \cref{eq:opt2} becomes:
\begin{equation*} 
\zeta_k := \max_{\omega, \alpha \in [0,1]} (1-\alpha) - \alpha \cdot \omega \qquad \mbox{s.t.} \qquad 
\beta \cdot \sum_{\ell = 1}^k {k \choose \ell} \alpha^{\ell} (1-\alpha)^{k-\ell} \cdot
\left[ \frac{\ell \cdot \omega}{\ell \cdot \omega + k-\ell } \right] + (1-\beta) \cdot \alpha \ge \frac{1}{2},
\end{equation*}
which can be optimized  as in \cref{sec:random_small} for any fixed $\beta$. As with the previous generalization, the distortion bound will be asymptotically larger than $1$ if $\beta < 1$.

%% file: conclusion.tex
\section{Open Questions}
We now list some open questions. At the technical level, the main question is to close the gap between the lower and upper bounds in \cref{tab:average,tab:random}. An intriguing question for $k=2$ is the following: If every pair of voters deliberates and outputs a ranking over alternatives that is consistent with the sum of their distances to the alternatives, is there a deterministic social choice rule over these rankings that beats distortion $3$? The lower bound example in \cref{thm:theta2} can be extended to weighted tournament rules~\cite{MunagalaW19,Kempe_2020}, and shows these rules are insufficient. Further, rules like plurality veto~\cite{Kizilkaya022} that output the favorite alternative of some voter are insufficient by \cref{eg1}. Finding a rule for $k=2$  that admits to tractable analysis, yet breaks the bound of $3$ is an interesting question.

More generally, tightening our bounds will require analysis of rules beyond tournaments, which poses analytic challenges. Even for tournament rules like Copeland, it would be interesting to bypass the use of non-convex optimization tools in deriving the results in \cref{sec:avg}, which will enable extensions to larger group sizes. Indeed, our conjecture is that  for all $k \ge 2$, the optimal solution to the non-convex program in \cref{sec:avg} has the same binary support as that of the lower bound instances in \cref{thm:lb1}. It would also be interesting to develop a tight distortion bound for Copeland for $k=3$, improving \cref{thm:theta3}. 

At the modeling level, we can consider richer models of how voters randomize between outcomes. For instance, the work of~\cite{goyal2025metricdistortionprobabilisticvoting}  proposes a randomized voting model where $\Pr[W \succ_i X] = g(r)$, where $r = d(i,W)/d(i,X)$, and $g(r)$ is a bounded, increasing function, such as $g(r) = \frac{r^2}{1+r^2}$. In this model, they show improved distortion bounds without deliberation, and it would be interesting to analyze extensions to deliberation. Further, in our model, we have assumed the groups are chosen by random sampling.  In practice, the sampling could be stratified based on voter features, or their prior ranking of the alternatives. It would be interesting to extend the model in \cref{sec:random} to handle such generalizations, and find the optimal stratification. Further, individuals often enter into deliberation with goals beyond simply persuading others, for instance, they may seek to understand different viewpoints, or strive for the common good~\cite{ashkinaze2024}. It would be interesting to model  these aspects formally. 

Finally, there are aspects of real-world deliberation that we have not modeled. For instance, party affiliation and signaling significantly impacts opinion formation and social learning (see~\cite{partisan1} and citations therein), since some people take their cues from a higher authority like party, or could defer to the partisan group, especially when they know little about a topic. 
Similarly, some individuals are better at persuasion than others (regardless of the opinions they hold), and it would be interesting to add  individual capability into the model.
